\documentclass{emulateapj-rtx4}
\usepackage{natbib}
\usepackage{epsf,graphics}
\usepackage{subfigure,graphicx}
\usepackage{bm}
\usepackage{amssymb}
\usepackage{amsbsy}
\usepackage{amsmath}
\usepackage{gensymb}
\usepackage{color}

\begin{document}


\title{An unexpected high concentration for the dark substructure in the gravitational lens SDSSJ0946+1006}

\author{Quinn E. Minor}
\affiliation{Department of Science, Borough of Manhattan Community College, 
City University of New York, New York, NY 10007, USA}
\affiliation{Department of Astrophysics, American Museum of Natural History, 
New York, NY 10024, USA}
\author{Sophia Gad-Nasr}
\author{Manoj Kaplinghat}
\affiliation{Department of Physics and Astronomy, University of California, 
Irvine CA 92697, USA}
\author{Simona Vegetti}
\affiliation{Max Planck Institute for Astrophysics, Karl-Schwarzschild-Strasse 1, D-85740 Garching, Germany}

\begin{abstract}
The presence of an invisible substructure has previously been detected in the gravitational lens galaxy SDSSJ0946+1006 through its perturbation of the lensed images. Using flexible models for the main halo and the subhalo perturbation to fit the lensed images, we demonstrate that the subhalo has an extraordinarily high central density and steep density slope. The inferred concentration for the subhalo is well above the expected scatter in concentrations for $\Lambda$CDM halos of similar mass. We robustly infer the subhalo's projected mass within 1 kpc to be $\sim 2$-$3.7\times 10^9$M$_\odot$ at $>$95\% CL for all our lens models, while the average slope of the subhalo's projected density profile over the radial range 0.75-1.25 kpc is constrained to be steeper than isothermal ($\gamma_{2D} \lesssim -1$). By modeling the subhalo light directly, we infer a conservative upper bound on its luminosity $L_V < 1.2\times 10^8L_\odot$ at 95\% CL, which shows that the perturber is dark matter dominated. To compare to $\Lambda$CDM expectations, we analyze subhalos within analogues of lensing galaxies in the Illustris TNG100-1 simulation over many lines of sight, and find hundreds of subhalos that achieve a projected mass within 1 kpc of $\gtrsim 2\times10^9M_\odot$. However, less than 1\% of the mock observations yield a log-slope $\gamma_{2D}$ steep enough to be consistent with our lensing models, and they \emph{all} have stellar masses in excess of that allowed by observations by about an order of magnitude or more. Comparison to the dark-matter-only TNG100-1 simulation suggests that these high projected masses and steep slopes are explained by adiabatic contraction due to their high stellar mass within 1 kpc ($\gtrsim 10^9M_\odot$), an insignificant effect for the observed subhalo given its low stellar mass. We conclude that the presence of such a dark, highly concentrated subhalo is unexpected in a $\Lambda$CDM universe. Finally, we show that this tension with CDM is not significantly reduced if the perturber is assumed to be a line-of-sight structure, rather than a subhalo.

\end{abstract}

\keywords{gravitational lensing: strong -- dark matter -- galaxies: dwarf\vspace{1.0mm}}

\section{Introduction}\label{sec:intro}

A key prediction of the Cold Dark Matter (CDM) paradigm is the presence of dark matter subhalos within larger host halos \citep{klypin1999,moore1999}. This prediction is shared by closely-related models like self-interacting dark matter and warm dark matter models. However, these models make distinct predictions for the density profile, concentrations and shapes of subhalos \citep{Vogelsberger:2012MNRAS.423.3740V,rocha2013,lovell2014}. Detecting these subhalos and inferring their properties provides an essential test of dark matter physics. 

In order to test the cold dark matter paradigm, one should ideally probe the structure of dark matter halos that are sufficiently ``dark'' such that baryonic physics cannot alter its overall mass or density profile significantly. There are two principle strategies to accomplish this. In the Milky Way, dark matter halos can be detected by their perturbations of stellar tidal streams \citep{carlberg2012,carlberg2013,erkal2015}. This strategy has recently led to the detection of a $10^6 - 10^8 M_\odot$ subhalo in the GD1 stellar stream \citep{bonaca2019}. In distant galaxies that act as strong gravitational lenses, dark matter subhalos (or small field halos along the line of sight) can be detected via their perturbations of highly magnified images \citep{mao1998,metcalf2001,koopmans2005}. The presence of dark matter substructure can be established statistically \citep{dalal2002,kochanek2004,xu2015,hezaveh2014,cyrracine2016}, or else individual perturbers can be detected for strong enough perturbations \citep{vegetti2009}; in the latter category, detections have been claimed in four lens systems thus far \citep{vegetti2010,vegetti2012,nierenberg2014,hezaveh2016}. A great many more subhalos are expected to be detected among the avalanche of strong lenses expected from the upcoming Euclid and LSST surveys, after high-resolution follow-up imaging of these lenses.

Several papers have compared the mass function of detected substructures (as well as field halos along the line of sight; cf. \citealt{despali2016,li2017}) perturbing lensed arcs to the expectation of CDM, showing broad consistency albeit with low statistical significance \citep{vegetti2014b,ritondale2019}. More recently, the mass-concentration relation of substructure in quasar lenses has been compared to the expectation of CDM \citep{gilman2020}. These approaches show great promise, but are still hampered by small number statistics. Additionally, the inferred dark matter halo masses themselves depend on assumptions about the density profiles and tidal radii of the subhalos, complicating the analysis \citep{minor2017}.

Another approach, completely unexplored to date, is to constrain the density profile of individual perturbers and test their inferred properties (e.g. concentration) against the expectation from $\Lambda$CDM \citep{vegetti2014}. While the concentration of relatively low-mass perturbers is difficult to infer, in principle the mass and concentration of large perturbers can be constrained. Despite only four individual detections, here we benefit from the fact that the largest perturbations are the most likely to be detected, and thus our initial set of detections is likely to be biased toward relatively massive and/or concentrated perturbers. The question is then, can we constrain the individual concentrations and masses of detected subhalos and test these against the expected mass-concentration relation and its scatter in CDM simulations?

In this paper we constrain, for the first time, the concentration of a dark matter subhalo perturbing a gravitationally lensed arc, originally discovered by \cite{vegetti2010} using HST observations of the lens system SDSSJ0946+1006. By comparing to analogous subhalos in the IllustrisTNG simulations, we will show that its remarkably high mass and concentration are in tension with CDM at the $>99\%$ confidence level. Moreover, the lack of a distinct stellar light signal (implying $L_V \lesssim 10^8 L_\odot$) implies that star formation and gas physics are unlikely to resolve the discrepancy. This motivates considering modified dark matter physics as a possible explanation.

The paper is organized as follows. In Section \ref{sec:lensmodeling} we will describe the lens modeling procedure. In Section \ref{sec:results} we describe the analysis carried out on SDSSJ0946+1006. The best-fit models are compared in Section \ref{sec:bestfit}, and the subhalo mass/concentration constraints are presented and compared to CDM predictions in Section \ref{sec:mc_constraints}. More robust constraints on the density profile---specifically a mass scale and density log-slope---are inferred in Sections \ref{sec:rpert} and \ref{sec:logslope} respectively. In Section \ref{sec:lv_constraint} we infer a conservative upper bound on the stellar luminosity of the subhalo. In Section \ref{sec:illustris} we compare these results to subhalos of analogous lens galaxies within the IllustrisTNG simulation to investigate whether CDM is capable of producing such high concentrations. The simulated sample is described in Section \ref{sec:illustris_sample}, while the comparison of subhalo candidates to the lensing constraints is made in Section \ref{sec:illustris_results}. In Section \ref{sec:illustris_slope_pct} we estimate the likelihood of such a high concentration subhalo generating a similar-sized subhalo perturbation in CDM, and interpret the results in Section \ref{sec:discussion}.  In Section \ref{sec:los} we discuss whether the tension with CDM can be reduced if the perturber is a field halo along the line-of-sight, rather than a subhalo. In Section \ref{sec:sidm} we discuss whether dark matter physics, in particular SIDM, can explain the high concentration of the subhalo. Finally, we give our conclusions in Section \ref{sec:conclusions}.

\section{Lens modeling}\label{sec:lensmodeling}

The gravitational lens SDSSJ0946+1006 is remarkable in that it is an individual galaxy that is lensing two different source galaxies at different redshifts, $z=0.609$ and $z\approx 2.4$ \citep{gavazzi2008,sonnenfeld2012,collett2014}, earning it the nickname ``the Jackpot''. (More recently a third, faint source at $z\approx 6$ has been found by \citealt{collett2020}). The subhalo discovered in \cite{vegetti2010} was found perturbing the images from the closest source, at $z=0.609$. Although stronger constraints on the mass distribution of the lens (and indirectly, the subhalo) may be possible by analyzing both lensed sources, this complicates the analysis and we therefore focus only on the lower redshift source galaxy in this paper.

Our method for modeling the data uses a combination of parameterized surface brightness profiles to represent the $z=0.609$ source galaxy. When evaluating the likelihood, image pixels are ray traced to the source plane and the following image is convolved with the PSF (details on the ray tracing are given at the end of this section). This allows for fast likelihood evaluations and rapid exploration of the parameter space, which consists of both 
lens and source model parameters, thereby reducing computational cost compared to reconstructing pixellated sources. However, to use elliptical profiles would be too restrictive a prior, for two reasons: 1) actual galaxies rarely (if ever) have perfectly elliptical isophotes; 2) even if the true source is elliptical, if one is modeling a subhalo with an incorrect mass or concentration, this systematic can often be (at least partially) absorbed into the inferred source galaxy by perturbing the isophotes. Thus, if one does not allow enough freedom in the source galaxy model, the danger arises that the constraints on concentration and mass may appear stronger than they really are.

To allow the requisite freedom in the source galaxy model, we start with a cored Sersic profile, where the core is defined by the replacement $r^2 \rightarrow r^2 + r_c^2$. We then perturb the isophotes to allow for non-elliptical profiles, as follows: first, the unperturbed isophotes are ``generalized ellipses'' with a radial coordinate defined by

\begin{equation}
    r_0(x,y) = \left(|x-x_0|^{C_0+2} + \left|\frac{y-y_0}{q}\right|^{C_0+2}\right)^\frac{1}{C_0+2}
\end{equation}

where $q$ is the axis ratio and $C_0$ is the ``boxiness'' parameter, such that $C_0=0$ corresponds to a perfectly elliptical profile. We then add Fourier mode perturbations to the isophotes, as follows:

\begin{equation}
    r(x,y) = r_0(x,y)\left\{1+\sum_{m=1}^N \left[a_m \cos(m\theta) + b_m\sin(m\theta)\right]\right\}
\end{equation}

From experimentation we find that parameter exploration can become difficult beyond 4-5 Fourier modes, and the $m=2$ mode is quite degenerate with the axis ratio parameter $q$. Thus, when modeling the diffuse components of the source galaxy, we include the modes $m=1,3,4,5,6$, with sine and cosine terms for each. For high $m$ modes, fluctuations can easily become quite rapid, leading to noisy source solutions. To regulate this, we switch to the scaled amplitudes $\alpha_m = m a_m$, $\beta_m = m b_m$, which are the amplitudes of the azimuthal derivative $dr/d\theta$. By using these scaled amplitudes as free parameters, we can set an upper prior limit on the rate of change of the contours that applies equally to all modes. Our method for perturbing the isophotes is essentially identical to that employed by the GALFIT algorithm for fitting galaxy images \citep{peng2010}, except that instead of including a phase angle parameter, we use the scaled amplitudes for both sine and cosine terms as free parameters; this ensures that there is no coordinate singularity in the limit of very small amplitudes. 

Since the Hubble PSF is relatively undersampled compared to the pixel size, the  images produced can be sensitive to how the ray tracing is done, particularly  near the critical curve. To mitigate this, we split each image pixel into  2$\times$2 subpixels and ray-trace the center point of each subpixel to the  source plane, assigning it the surface brightness given by the source profile  at the position of the ray-traced point. These surface brightness values are  then averaged to find the surface brightness of the image pixel.  After the  ray-tracing is complete, the resulting pixel values are then convolved with the  PSF.  We find that 2$\times$2 splitting achieves sufficient accuracy without adding too much computational burden; a greater number of subpixels changes the surface brightness values very little while adding significant computational cost.

In a companion paper (Minor et al., in prep) we apply this method of lens modeling to a large number of mock data with a gravitational lens similar to that of SDSSJ0946+1006, but with a wide variety of subhalo perturbations. We find that for sufficiently large perturbations, the mass and concentration can be meaningfully constrained, and are recovered without bias. However, the inferred mass and concentration of the perturbing subhalo are degenerate for sufficiently small perturbations. A more direct constraint is provided by inferring the subhalo's projected mass within a characteristic radius and the average slope near this radius as derived parameters, which are also well recovered for perturbations in the mock data (Minor et al. in prep, Figure 6), justifying the approach used in this paper. These derived parameters will be motivated in Sections \ref{sec:rpert} and \ref{sec:logslope}.

\section{Results of lensing analysis}\label{sec:results}

\begin{table*}[t]
\centering
\begin{tabular}{|l|c|c|c|c|l|}
\hline
 & tNFW & tNFWmult & CoreCusp & CoreCuspmult & Prior\\
\hline
$R_e$(arcsec) & $1.362_{-0.002}^{+0.002}$ & $1.347_{-0.005}^{+0.005}$ & $1.363_{-0.001}^{+0.001}$ & $1.359_{-0.002}^{+0.002}$
 & [1.3,1.4] \\
$\alpha$ & $1.32_{-0.04}^{+0.04}$ & $1.37_{-0.05}^{+0.05}$ & $1.34_{-0.05}^{+0.05}$ & $1.30_{-0.05}^{+0.05}$
 & [0.5,1.5]\\
$e_1 \equiv (1-q)\cos(2\theta)$ & $0.057_{-0.009}^{+0.010}$ & $0.068_{-0.011}^{+0.011}$ & $0.061_{-0.013}^{+0.014}$ & $0.056_{-0.010}^{+0.010}$
 & [-0.15,0.15]\\
$e_2 \equiv (1-q)\sin(2\theta)$ & $-0.041_{-0.009}^{+0.009}$ & $-0.034_{-0.009}^{+0.010}$ & $-0.050_{-0.009}^{+0.009}$ & $-0.032_{-0.009}^{+0.010}$
 & [-0.15,0.15]\\
$x_c$(arcsec) & $0.0042_{-0.0020}^{+0.0024}$ & $0.0132_{-0.0027}^{+0.0028}$ & $0.0041_{-0.0019}^{+0.0017}$ & $0.0102_{-0.0026}^{+0.0026}$
 & [-0.05,0.05]\\
$y_c$(arcsec) & $-0.0016_{-0.0020}^{+0.0026}$ & $-0.0162_{-0.0032}^{+0.0035}$ & $-0.0018_{-0.0024}^{+0.0024}$ & $-0.0100_{-0.0028}^{+0.0029}$
 & [-0.05,0.05]\\
$\Gamma_1$  & $0.074_{-0.003}^{+0.003}$ & $0.072_{-0.003}^{+0.003}$ & $0.075_{-0.004}^{+0.004}$ & $0.070_{-0.003}^{+0.003}$
 & [-0.1,0.1]\\
$\Gamma_2$ & $-0.069_{-0.002}^{+0.002}$ & $-0.071_{-0.003}^{+0.003}$ & $-0.070_{-0.003}^{+0.003}$ & $-0.070_{-0.003}^{+0.003}$
 & [-0.1,0.1]\\
$A_3$ & ... & $0.0067_{-0.0020}^{+0.0021}$ & ... & $0.0067_{-0.0021}^{+0.0019}$
 & [-0.04,0.04]\\
$B_3$ & ... & $-0.0174_{-0.0031}^{+0.0030}$ & ... & $-0.0146_{-0.0029}^{+0.0033}$
 & [-0.04,0.04]\\
$A_4$ & ... & $0.0174_{-0.0026}^{+0.0027}$ & ... & $0.0169_{-0.0027}^{+0.0028}$
 & [-0.04,0.04]\\
$B_4$ & ... & $-0.0228_{-0.0038}^{+0.0038}$ & ... & $-0.0195_{-0.0040}^{+0.0037}$ & [-0.04,0.04]\\
\hline
$\log(m_{200}/M_\odot)$ & $9.70_{-0.09}^{+0.11}$ & $10.48_{-0.13}^{+0.12}$ & ... & ... & [8,11.95]\\
$r_{s}(kpc)$ & $0.021_{-0.011}^{+0.029}$ & $0.844_{-0.230}^{+0.260}$ & ... & ... & [0.01,5], log\\
$\log(\kappa_0)$ & ... & ... & $0.184_{-0.332}^{+0.113}$ & $0.096_{-0.239}^{+0.149}$ & [-2.0,0.3]\\
$\gamma_{inner}$ & ... & ... & $1.742_{-0.211}^{+0.302}$ & $0.766_{-0.251}^{+0.379}$ & [0.5,2.9]\\
$r_{t}(kpc)$ & $24.2_{-19.9}^{+68.5}$ & $38.7_{-25.7}^{+56.8}$ & $0.586_{-0.056}^{+0.197}$ & $1.288_{-0.204}^{+0.351}$
 & [0.1,100], log\\
$x_{c,sub}$(arcsec) & $-0.652_{-0.010}^{+0.008}$ & $-0.691_{-0.018}^{+0.016}$ & $-0.646_{-0.011}^{+0.011}$ & $-0.698_{-0.020}^{+0.017}$
 & [-1.7,1.2]\\
$y_{c,sub}$(arcsec) & $0.950_{-0.019}^{+0.011}$ & $1.009_{-0.024}^{+0.026}$ & $0.956_{-0.014}^{+0.019}$ & $1.010_{-0.021}^{+0.022}$
 & [-1.8,1.5]\\
\hline
$M_{2D}$(1kpc)$(10^9M_\odot)$ & $2.50_{-0.31}^{+0.29}$ & $3.33_{-0.26}^{+0.3}$ & $2.75_{-0.28}^{+0.24}$ & $3.31_{-0.28}^{+0.31}$
 & ...\\
$\gamma_{2D}(0.75,1.25$ kpc$)$ & $-1.98_{-0.16}^{+0.05}$ & $-1.27_{-0.13}^{+0.11}$ & $-3.27_{-0.17}^{+0.27}$ & $-1.79_{-0.25}^{+0.23}$
 & ...\\
$m_{sub,tot}(10^{10}M_\odot)$ & $0.469_{-0.154}^{+0.202}$ & $2.61_{-1.06}^{+1.29}$ & $0.312_{-0.040}^{+0.035}$ & $0.700_{-0.113}^{+0.131}$
 & ...\\
$c_{200}$ & $1560_{-859}^{+1440}$ & $70.5_{-11.9}^{+17.8}$ & ... & ... & ...\\
$r_{\delta c}$(arcsec) & $0.351_{-0.013}^{+0.010}$ & $0.308_{-0.014}^{+0.016}$ & $0.355_{-0.011}^{+0.009}$ & $0.323_{-0.016}^{+0.0160}$
 & ...\\
$\frac{M(r_{\delta c})}{\alpha}(10^9M_\odot)$ & $2.02_{-0.26}^{+0.25}$ & $2.69_{-0.25}^{+0.31}$ & $2.14_{-0.22}^{+0.19}$ & $2.91_{-0.30}^{+0.33}$
 & ...\\
\hline
\hline
$\log_{10}\mathcal{E}$ & -1537.7 & -1504.9 & -1537.7 & -1510.0 & ...\\
$\chi^2_{bf}$/pixel & 1.66 & 1.61 & 1.66 &   1.63 & ...\\
\hline
\end{tabular}
\caption{Inferred lens model parameters and their priors, and log-evidence and best-fit $\chi^2$ per pixel. The first 12 parameters describe the primary lens galaxy, while the following 5 parameters describe the subhalo. The ellipticity components are defined in terms of the axis ratio $q$ and orientation $\theta$ (measured east of north) as shown in the first column. The final 6 parameters are derived parameters that relate to the subhalo, namely the projected mass enclosed within 1 kpc, the average (2D) log-slope of the projected density profile from 0.5 kpc to 1kpc; the total subhalo mass; the concentration $c_{200}$; the perturbation radius; and the mass enclosed within the perturbation radius divided by the log-slope $\alpha$ of the primary lens galaxy. The uncertainties are Bayesian credible intervals derived from the 2.5\% and 97.5\% percentiles of the posterior probability distributions, which are equivalent to 2$\sigma$ errors if the distribution is Gaussian. Note that we omit the inferred non-linear parameters related to the source galaxy, which is visualized in Figure \ref{fig:source}.}
\label{tab:models}
\end{table*}

\begin{figure*}
	\centering
	\subfigure[masked HST image]
	{
		\includegraphics[height=0.32\hsize,width=0.31\hsize]{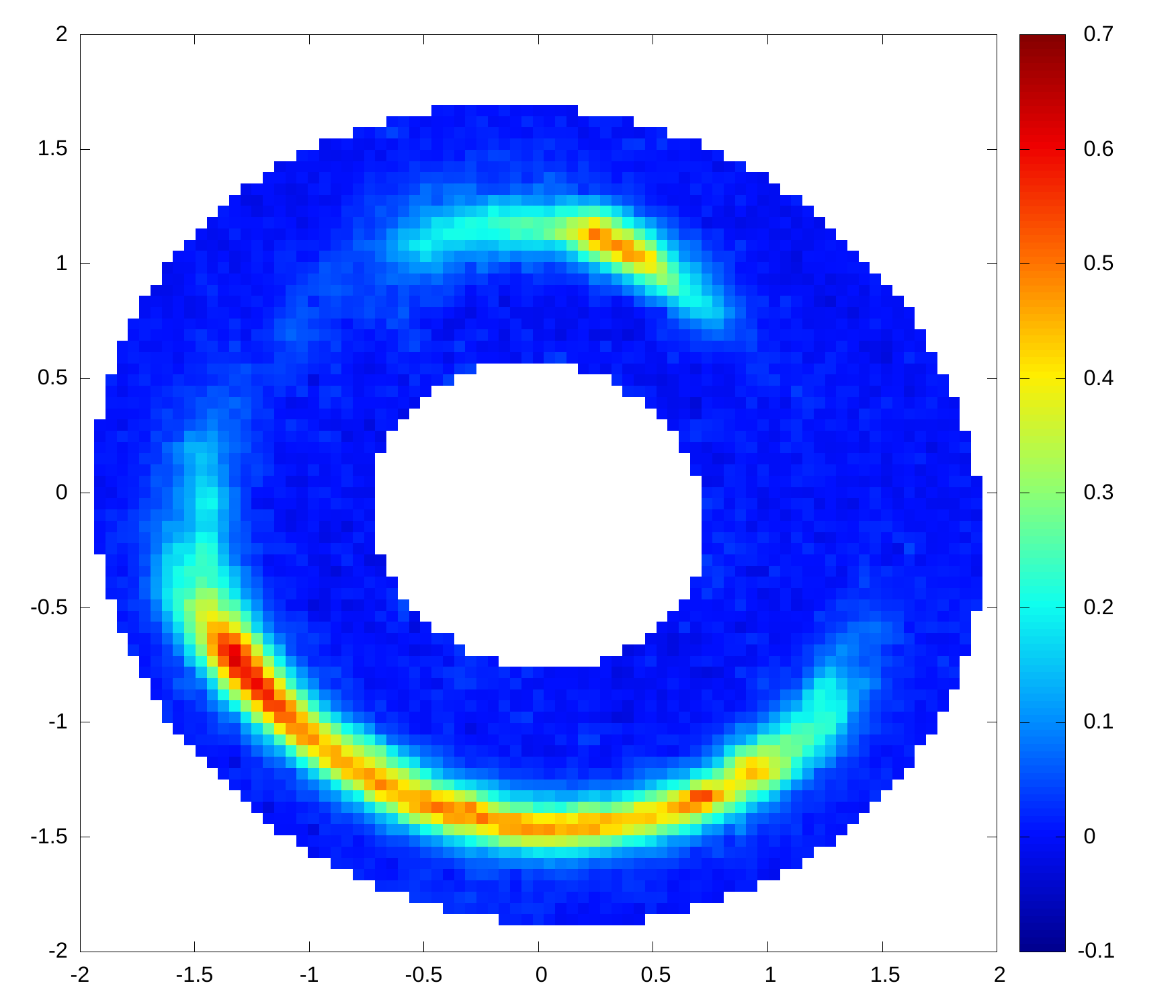}
		\label{dataimg}
	}
	\subfigure[best-fit model, tNFW]
	{
		\includegraphics[height=0.32\hsize,width=0.31\hsize]{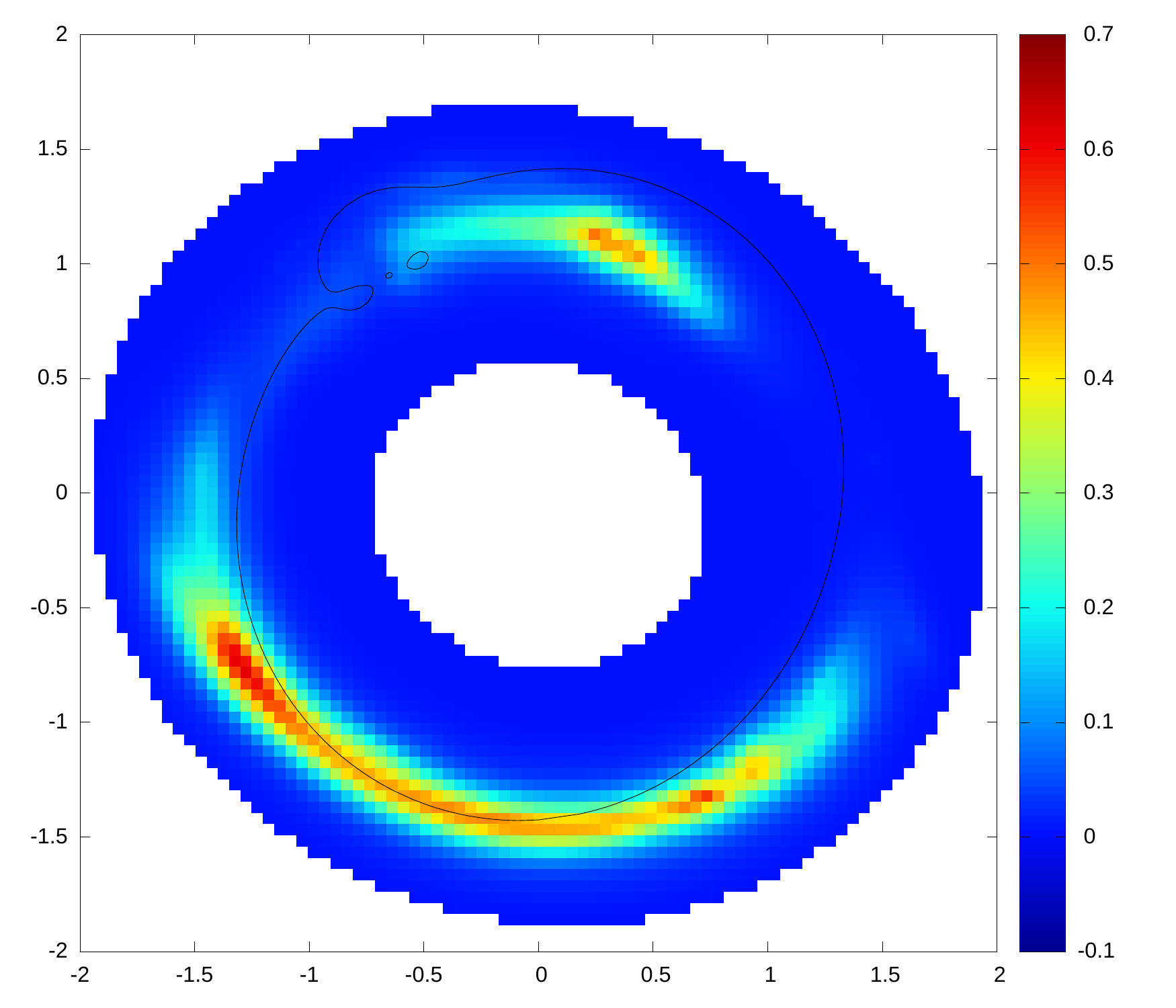}
		\label{modelimg1}
	}
		\subfigure[residuals, tNFW]
	{
		\includegraphics[height=0.32\hsize,width=0.31\hsize]{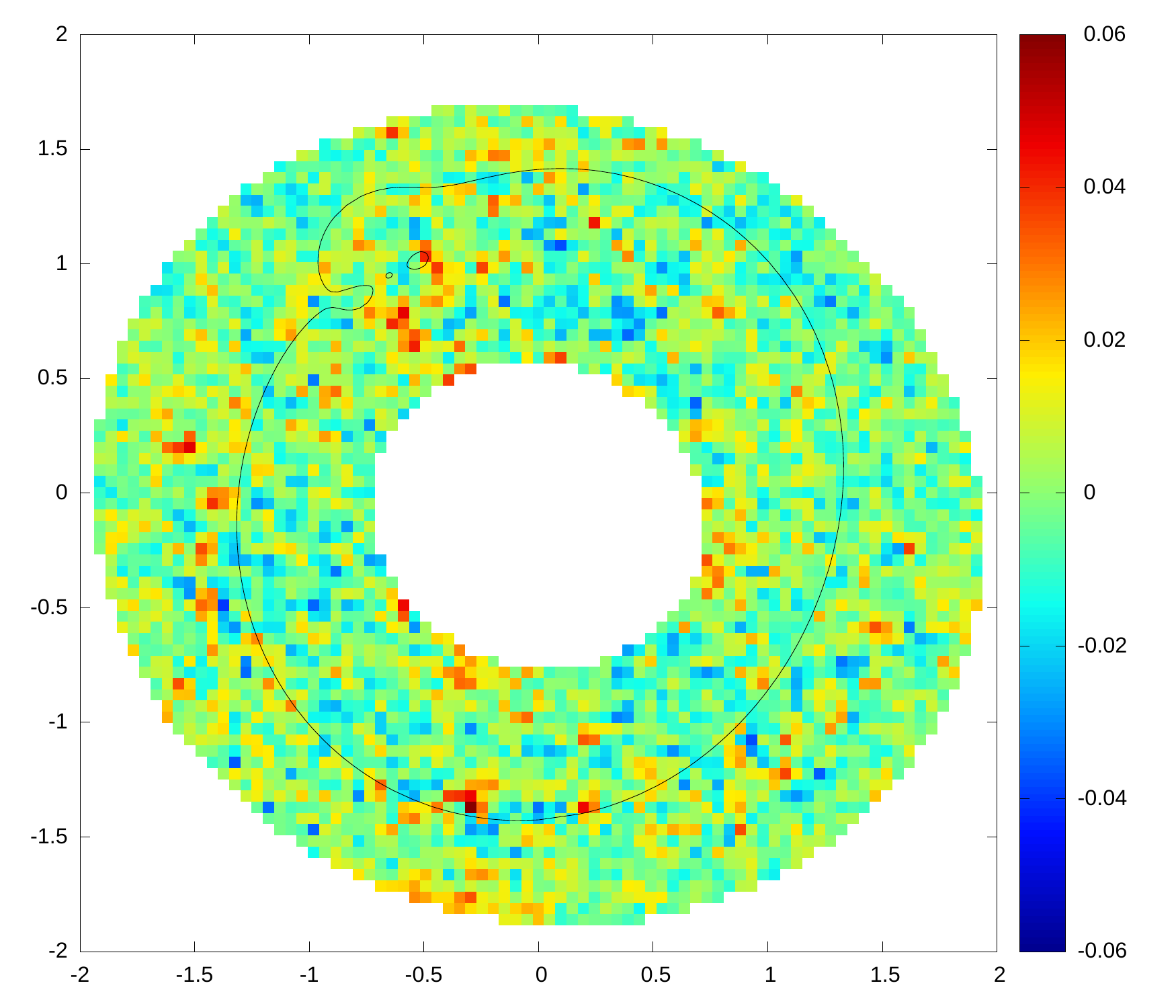}
		\label{residuals1}
	}
		\subfigure[best-fit model, tNFWmult]
	{
		\includegraphics[height=0.32\hsize,width=0.31\hsize]{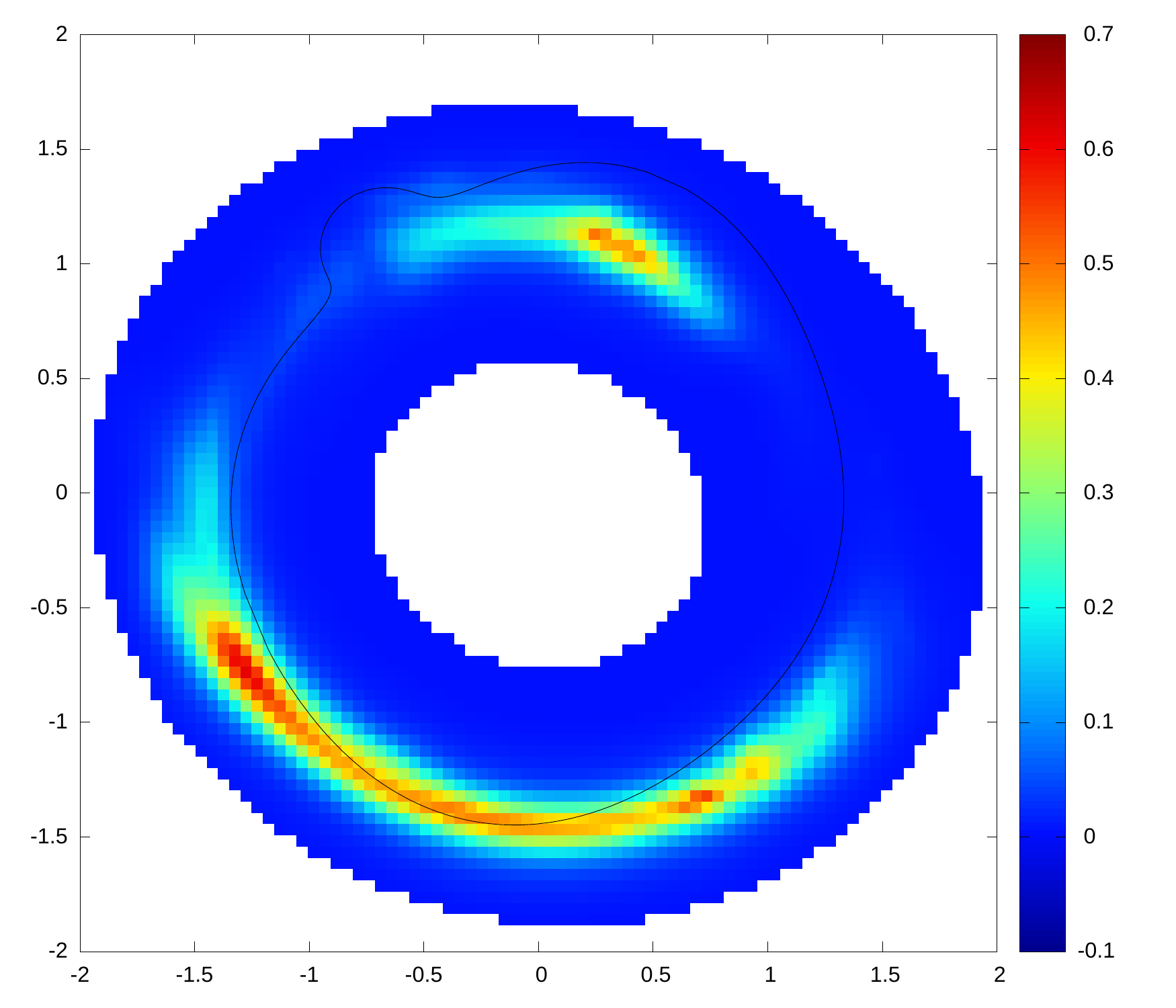}
		\label{modelimg2}
	}
		\subfigure[residuals, tNFWmult]
	{
		\includegraphics[height=0.32\hsize,width=0.31\hsize]{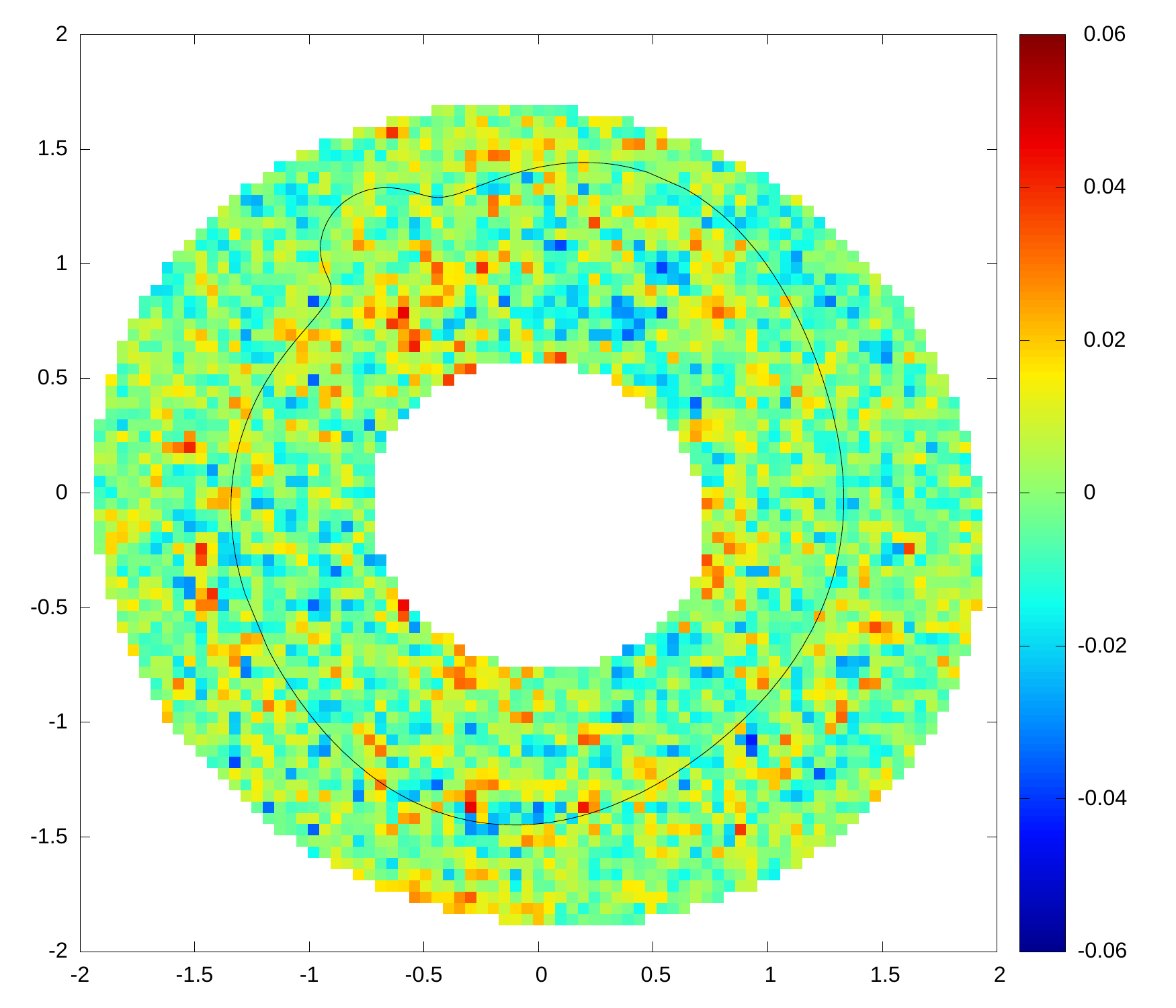}
		\label{residuals2}
	}
	\caption{HST image along with our best-fit lensing reconstructions and residuals using a truncated NFW profile for the subhalo. In (a) we show the  foreground-subtracted HST F814W image from \cite{gavazzi2008}. The best-fit model lensed images and residuals for the ``tNFW'' model with elliptical host galaxy are shown in (b) and (c). Figures (d) and (e) show the same for our ``tNFWmult'' model where multipoles are added to the host galaxy's projected density; this is the model most favored by the Bayesian evidence. The black curves are the critical curves of the lens mapping, which show the strong perturbation by the subhalo in the upper left.}
\label{fig:bestfit}
\end{figure*}

\begin{figure*}
	\centering

	\subfigure[reconstructed source, analytic]
	{
		\includegraphics[height=0.38\hsize,width=0.37\hsize]{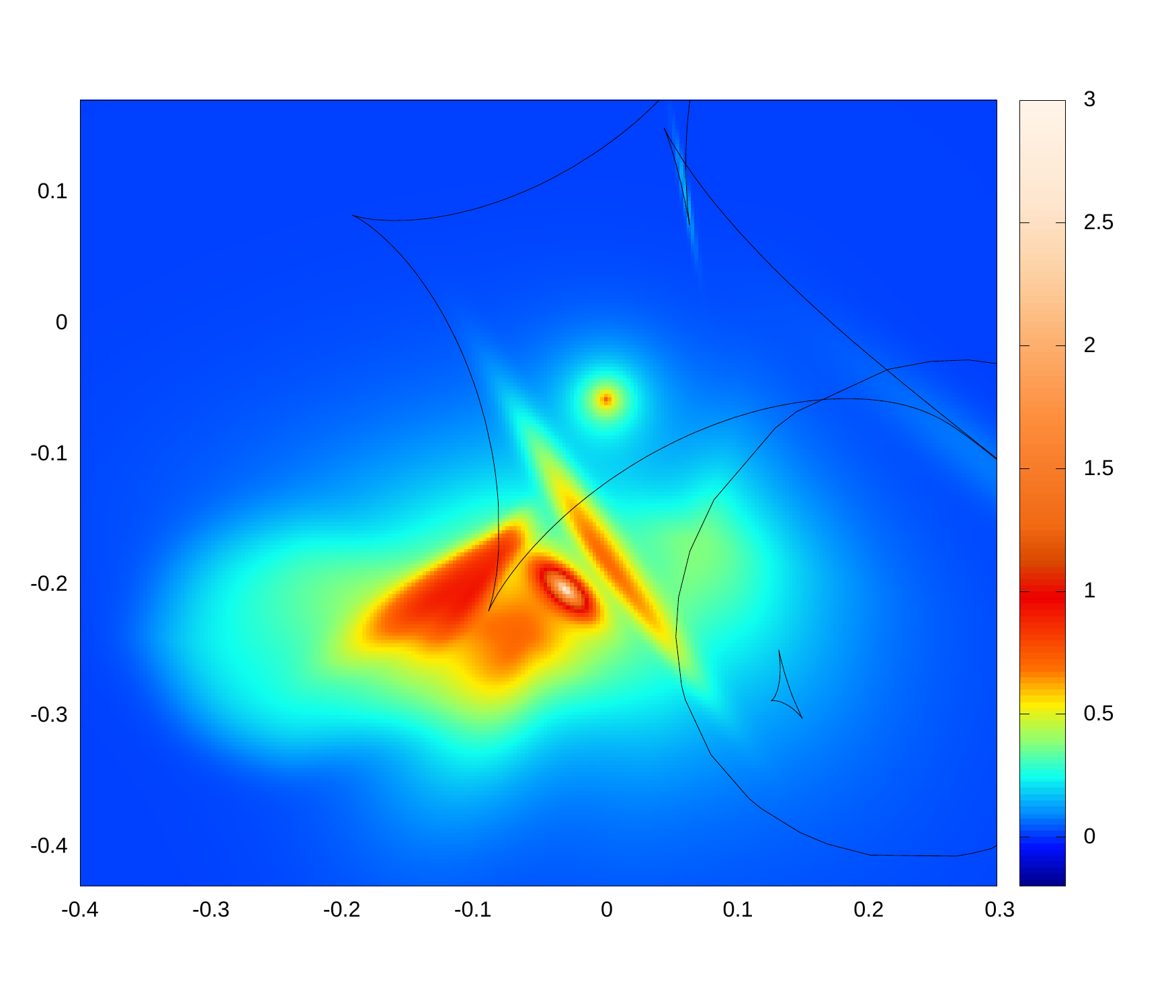}
		\label{modelsrc1}
	}
	\subfigure[reconstructed source, pixellated]
	{
		\includegraphics[height=0.38\hsize,width=0.37\hsize]{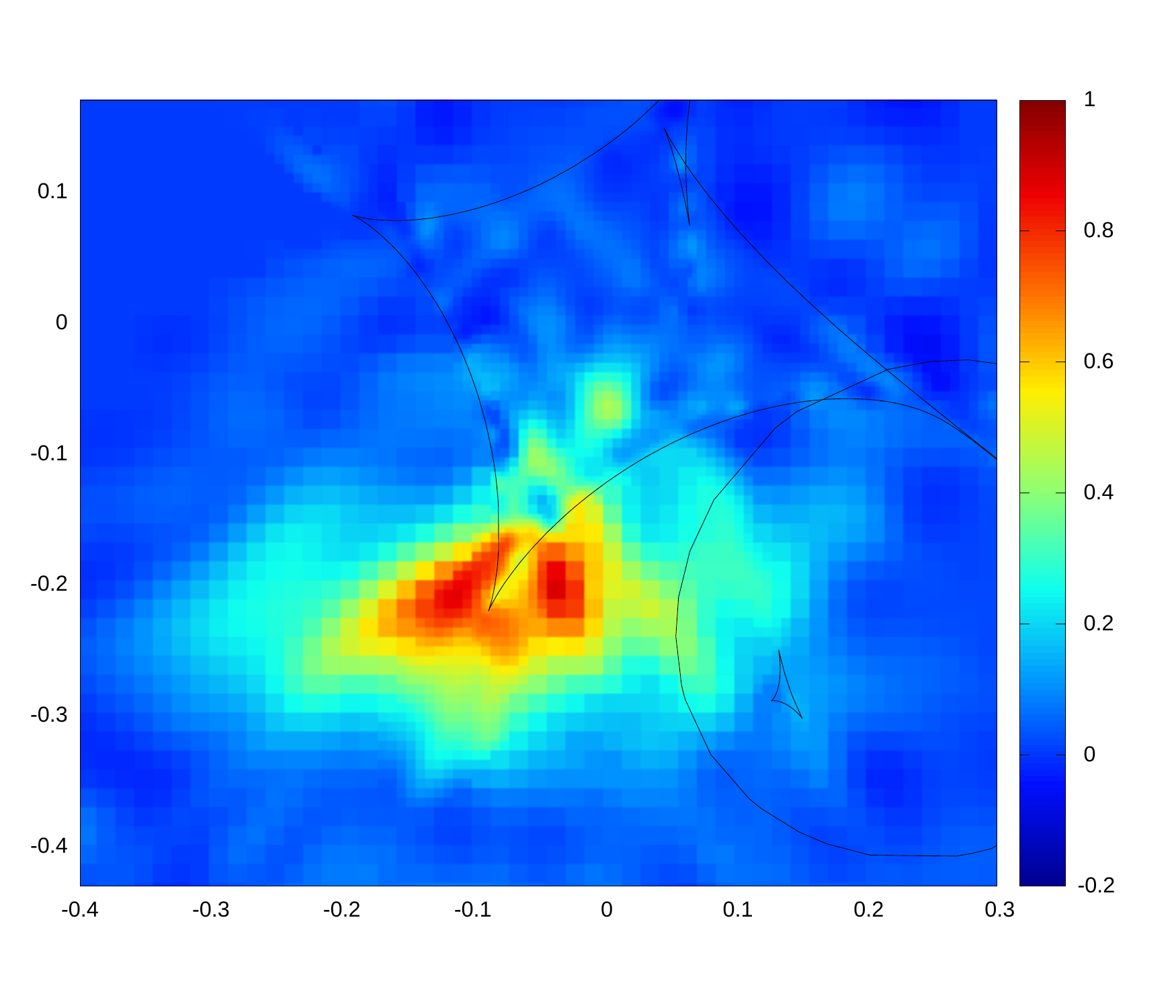}
		\label{modelsrc2}
	}
	\caption{Reconstructed source galaxy, analytic versus a pixel inversion from the best-fit tNFW model with elliptical host galaxy. The analytic source contains a region with relatively high surface brightness, however the color scale is chosen so that the colors in each figure correspond to approximately the same surface brightness values. Black curves show the caustic curves of the lens mapping; note the large caustic generated by the subhalo in the lower left.}
\label{fig:source}
\end{figure*}

For our primary modeling runs, the subhalo is modeled with a smoothly truncated Navarro-Frenk-White (tNFW) profile \citep{baltz2009} with outer log-slope of -5 well beyond the truncation radius (for comparison, a different subhalo profile will also be considered in Section \ref{sec:logslope}). All parameters of the density profile were varied freely: the mass $m_{200}$, scale radius $r_s$, and truncation radius $r_t$. Note that although we are describing a subhalo, one can formally define $m_{200}$ as the virial mass such a halo would have in the field in the absence of tidal stripping, thus providing an upper limit on the inferred subhalo mass. (The true mass of the subhalo will also depend on how it is truncated, meaning it is determined by both $m_{200}$ and $r_t$; note that even for a subhalo with very little tidal stripping, $r_t$ may indicate the virial radius at infall, which would be smaller than the virial radius at the lens redshift.) For ease of parameter exploration, for our mass parameter we use $\log_{10}(m_{200})$ and adopt a uniform prior in this parameter; we vary $r_s$ and $r_t$ but adopt log priors in these parameters. The prior limits on the lens parameters are listed in Table \ref{tab:models}.

We considered two different primary models. In model tNFW, the host lens galaxy is modeled with just an elliptical power law profile; in model tNFWmult, we also use an elliptical power law, but in addition we add $m=3$ and $m=4$ multipole terms to the projected density of the lens, with the same power law index as the elliptical component, to capture departures from ellipticity for the primary lens. The latter model is inspired by the fact that the lens galaxy light shows evidence of perturbed isophotes, possibly as a result of a recent encounter with a nearby galaxy \citep{sonnenfeld2012,gavazzi2008}. \cite{gavazzi2008} showed that approximately half of the mass within the Einstein radius is baryonic, hence the total projected density profile is likely to show marked deviations from ellipticity. This is further supported by the fact that \cite{vegetti2010} showed that their solution preferred to have smooth corrections to the gravitational potential of the lens (which they modeled as pixellated potential corrections) that indicated departures from ellipticity.

Although using analytic source profiles allows for very fast likelihood evaluations compared to pixellated source inversion, it has the disadvantage of being less automated in the sense that the user must choose how many profiles to employ, whether to add boxiness or Fourier modes, and what prior limits to choose. Thus, achieving a good fit requires at least a few iterations. For our initial run, we simply used one cored Sersic profile with a boxiness parameter and four Fourier modes ($m=1,3,4,5)$. Remarkably, although substantial residuals remained in our best-fit model, the lens parameters are roughly consistent with \cite{vegetti2010} and the subhalo was still detected in approximately the same location. This suggests that the existence and location of the subhalo is not strongly dependent on the source prior, although we will see that its location does vary slightly from model to model.

Likewise, doing a pixel inversion from our best-fit model yielded a source pixel map that is quite similar to the inferred source in \cite{vegetti2010} (Figure \ref{modelsrc2} shows our final inversion). From the pixel map, we recognize two bright regions near the center, as well as a possible peak with high ellipticity that intersects the lower-right part of the caustic; in the next iteration we add Sersic profiles to represent these, with the addition of two Fourier modes ($m=1,m=3$). In addition, there is also a small bright spot near the center of the caustic, so we add a spherical Sersic profile to capture this. It also became clear that the large diffuse component is not fully captured by a single profile, even with the addition of Fourier modes. In our final iteration, we thus use two cored Sersic profiles to represent the diffuse component, each with boxiness parameters and four Fourier modes ($m=1,3,4,5)$. In addition, there is fainter emission near the upper cusp of the astroid caustic, which is clearly perturbed by the subhalo, as well as within the additional caustic generated by the subhalo; finally, there is a large region of faint emission to the left of the large diffuse component. We add cored Sersic profiles to capture these, including two Fourier modes for the largest component. In total, then, we include eight source profiles in our fit, for a grand total of 109 parameters in the tNFW model, and 113 parameters in the tNFWmult model (with the addition of $m=3,m=4$ multipoles). The two profiles representing the primary diffuse component each have four Fourier modes, while the three largest additional components include only two ($m=1,3$) Fourier modes.

\begin{figure*}
	\centering
	\subfigure[tNFW model]
	{
		\includegraphics[height=0.32\hsize,width=0.34\hsize]{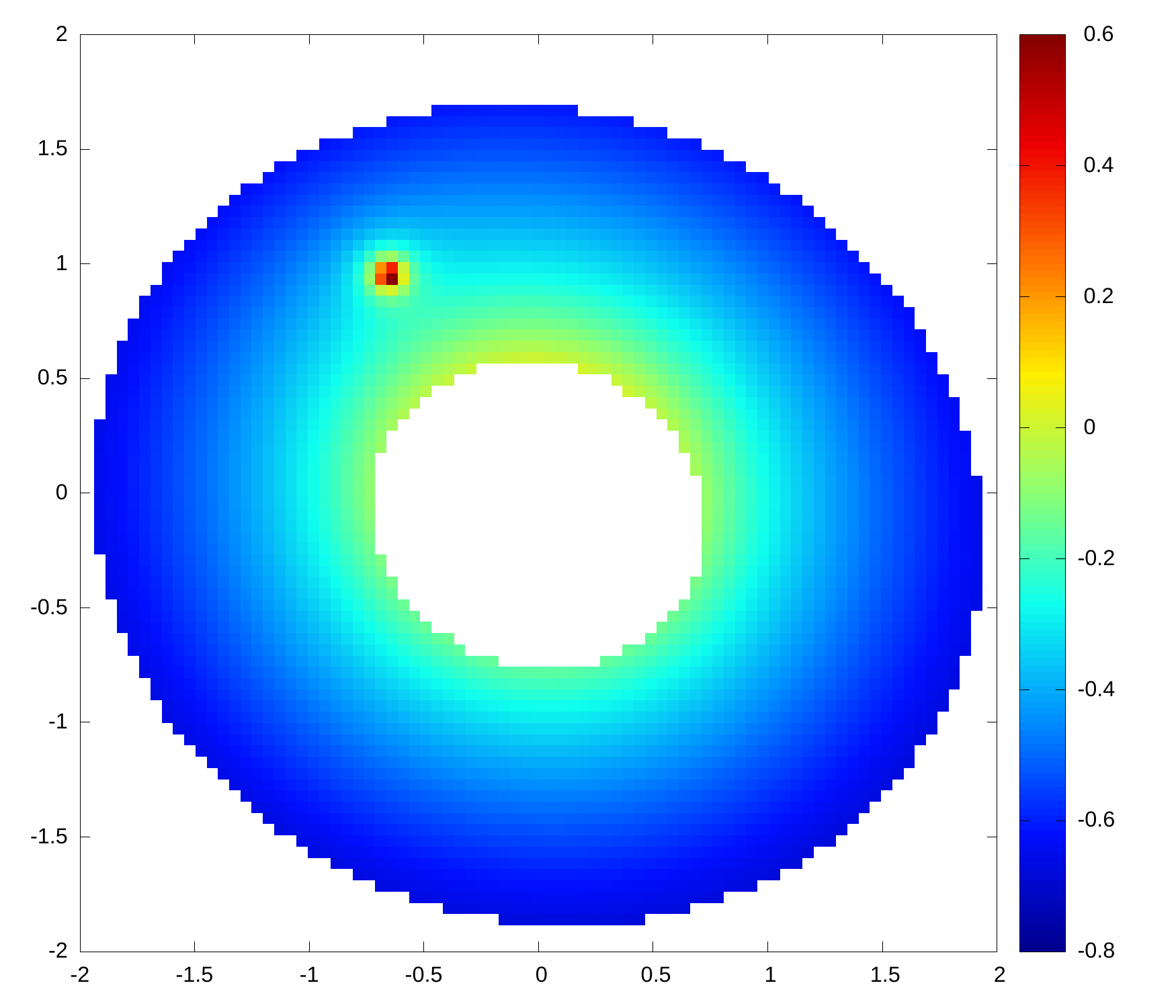}
		\label{lmodel1}
	}
	\subfigure[tNFWmult model]
	{
		\includegraphics[height=0.32\hsize,width=0.34\hsize]{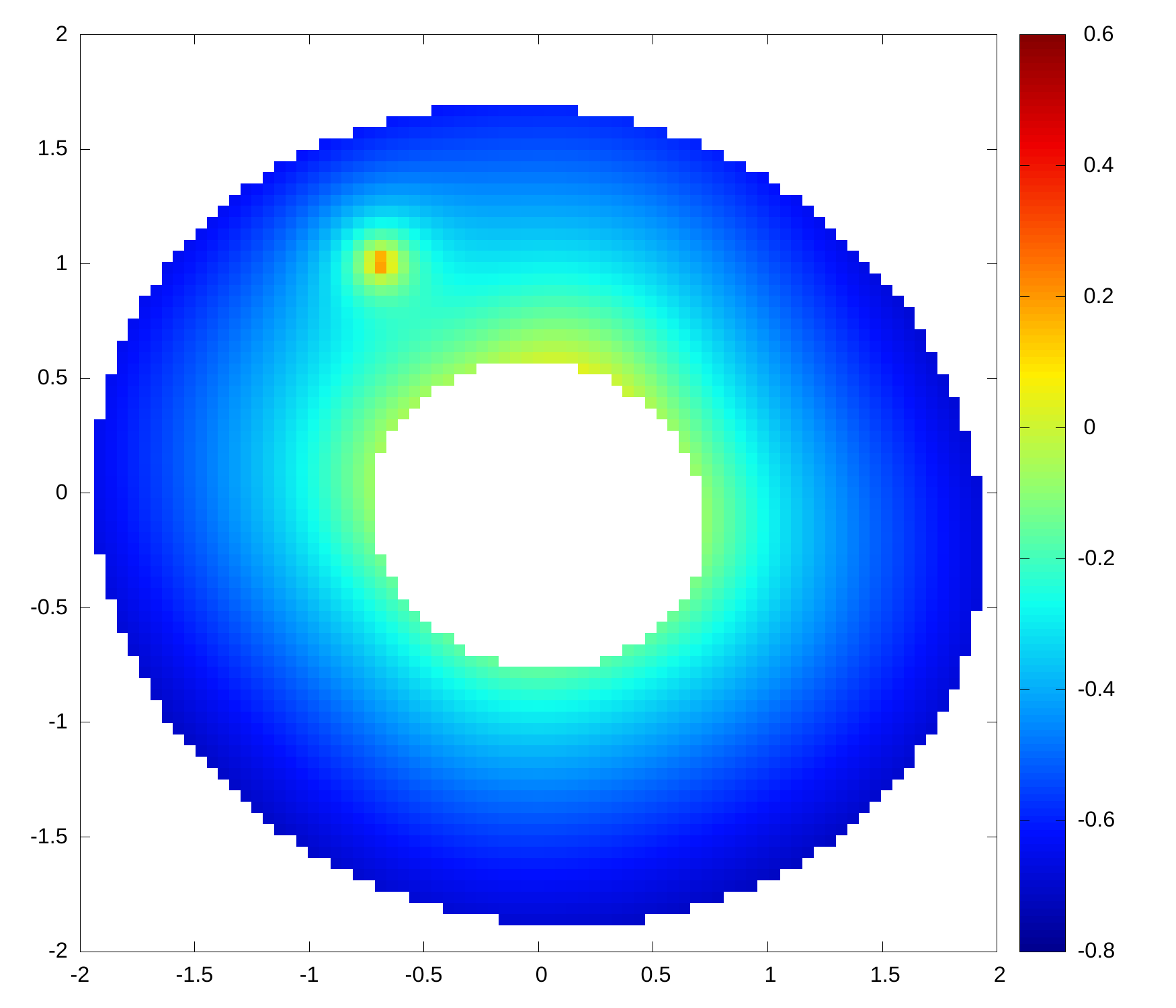}
		\label{lmodel2}
	}
	\caption{Projected density of the lens in each model, plotted as log(convergence) values at each image pixel within the mask. Note the contours are notably boxy in the tNFWmult model, which includes multipole terms in the projected density to capture departures from ellipticity and is preferred by the Bayesian evidence.}
\label{fig:logkappa}
\end{figure*}

The pixel mask we choose is identical to that of \cite{vegetti2010}, while we adopt the same model PSF as in \cite{vegetti2010,gavazzi2008}. To handle the large number of non-linear parameters in our final run, we use the PolyChord nested sampling algorithm, which evaluates the Bayesian evidence in addition to posterior samples.

\subsection{Best-fit models}\label{sec:bestfit}

In Figure \ref{fig:bestfit}, we plot the data image along with lensed images and residuals (with critical curves) from the best-fit parameters in models tNFW and tNFWmult, using the analytic source model outlined above. The location of the subhalo is quite evident from its strong perturbation of the critical curve in the upper left. The residuals are down to noise-level in many places, with very little correlated residuals along the location of the arcs, indicating our analytic model is capturing the primary features of the source galaxy. In Figure \ref{fig:source} we show the best-fit analytic source and caustics from the tNFW model (with tNFWmult being quite similar), and compare this to a source reconstruction obtained by a source pixel inversion from the same lens model. For the latter, we use a Cartesian grid with adaptive splitting \citep{dye2005} if the pixel magnification is 4 or greater, and curvature regularization where the regularization parameter is chosen to optimize the Bayesian evidence \citep{suyu2006}. The color bars are chosen such that identical colors in either figure represent the same surface brightness values. The comparison shows that the essential components are clearly well-reproduced by the analytic source, while fluctuations due to fitting noise are largely absent. Note that the analytic source has a bright region which peaks with a considerably higher surface brightness compared to the pixellated source; this is because such a high peak is discouraged in the pixel version due to the regularization prior which enforces smoothness even in the central regions of the source.

Close inspection of Figure \ref{fig:bestfit} shows that below the subhalo, there is some residual signal (up to $\approx4\sigma_{noise}$), which at first glance might seem to indicate a detail of the source that is not being well-captured. We found that placing a faint source blob near (0.4,-0.4) in the source plane can remove approximately half of the residual emission below the subhalo---however, this produces a partner image to the lower-right of the bottom arc, partly outside the mask, and with greater magnification. When the mask is removed, however, we found that there is no discernible signal where the partner image is supposed to be in the HST image. For the same reason, close inspection shows these residuals are not entirely eliminated in the pixel reconstructions in \cite{vegetti2010}. We therefore argue that the residual signal below the subhalo is likely due to unsubtracted foreground light from the primary lens galaxy. (This issue will be investigated further in Section \ref{sec:lv_constraint}.) The same systematic may also be responsible for other faint correlated residuals in the vicinity of the top arc. We note, however, that the presence of the subhalo is not sensitive to such faint foreground features---indeed we verified that the same subhalo solution is inferred even if the region surrounding the residual signal is masked out. We note that for the brightest pixels near the subhalo, containing the leftmost ``knob'' of the top arc, the residuals are consistent with the background noise (i.e. less than $3\times\sigma_{noise} \approx 0.01$).

The best-fit parameters and uncertainties are listed in Table \ref{tab:models}, along with the Bayesian model evidence and best-fit $\chi^2$/pixel values (the latter are calculated after doing a final optimization from the best-fit point in the chain using Powell's method). For reference, joint posteriors in the subhalo parameters for the tNFWmult model are plotted in Figure \ref{fig:posts} of Appendix \ref{sec:appendix_lensmodels}. The Einstein radius of the primary lens is $\approx 1.35-1.36''$, depending on the model, which is slightly higher than in \cite{vegetti2010} but significantly lower the $1.4''$ value inferred by \cite{collett2014}; likewise, the latter infer a slope that is near-isothermal ($\alpha\approx1.03$), compared to $\approx 1.3$ in our analysis. This may be explained by the covariance between the subhalo mass and the Einstein radius, as seen in Figure \ref{fig:posts} : as the subhalo mass is reduced, the Einstein radius must be increased. Hence, in the limit of zero subhalo mass, a significantly larger Einstein radius would be required to fit the data. Likewise, our posteriors show that a larger Einstein radius is correlated with a shallower slope, which may explain the difference in the inferred slope parameter.

Between the different subhalo models, the Bayesian evidence strongly prefers the tNFWmult model, with an evidence ratio of nearly 30 orders of magnitude. Indeed, close inspection of Figure \ref{fig:bestfit} reveals that certain residuals are improved in the tNFWmult model; this should be taken with some caution however, since it is possible (in principle) that lending more freedom to the source model might allow tNFW to achieve nearly as a good a fit as in tNFWmult. Nevertheless, there are certain residuals that are also present in the best-fit models in \cite{vegetti2010} which used pixellated sources, and at least a few of these residuals (notably above the center region of the bottom arc) do seem to be somewhat improved by the shifting of critical curves achieved in the tNFWmult model.

To get a better sense of how the solutions differ, in Figure \ref{fig:logkappa} we show the projected density of each best-fit solution by plotting the log(convergence) values at each pixel within the mask. Note that in tNFWmult the contours are notably boxy as a result of the high multipole amplitudes $A_4, B_4$, whereas the subhalo is less concentrated (although more massive, as we will see shortly). By comparison, the galaxy isophotes do show departures from ellipticity, but not nearly as boxy as the density contours in tNFWmult. It should be emphasized, however, that the data are not sensitive to the density contours everywhere; there are no images on the right-hand side of the mask, for example, and thus no direct information on the shape of the contours there. Thus, while this solution may indicate departures from ellipticity in the vicinity of the images, it does not necessarily imply the \emph{complete} contours have the boxy shape indicated by Figure \ref{fig:logkappa}. Repeating our analysis with a pixellated source may shed further light on the question of whether the data prefers such a strong departure from ellipticity. The important takeaway here is that the way we model the shape of the primary galaxy's contours can influence the properties of the inferred subhalo, a feature we will discuss further in Section \ref{sec:rpert}. 

\begin{figure}
	\centering
	\includegraphics[height=0.92\hsize]{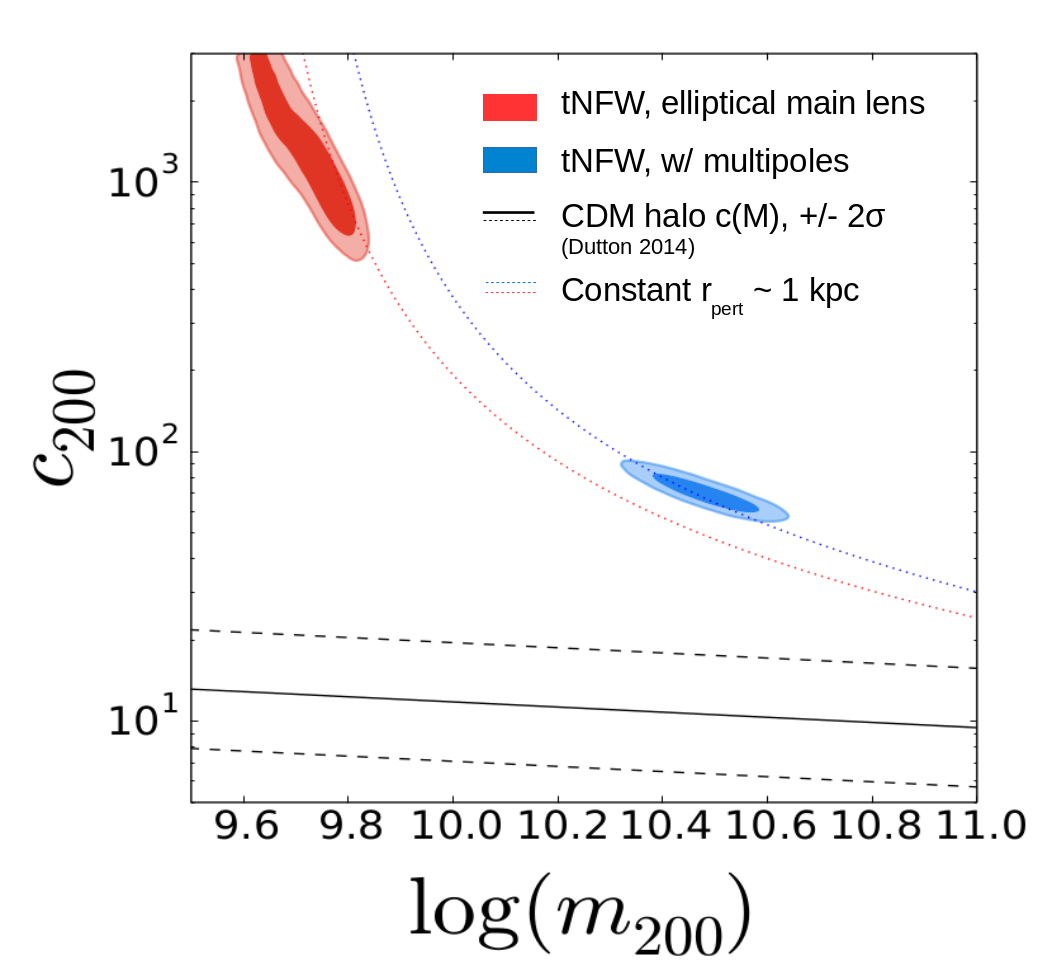}
	\caption{Posteriors in log($m_{200}$) versus concentration $c_{200}$ for the two fiducial models. For each model, the dark/light shaded regions enclose 68\% and 95\% of the total probability, respectively. Black solid curve denotes the median halo concentration-mass relation, along with $\pm 2\sigma$ scatter (black dashed curves). Curves of constant perturbation radius are shown (blue/red curves), generated by starting from the best-fit models, varying $m_{200}$, and finding the concentration that reproduces the mass within the perturbation radius (approximately 1 kpc).}
\label{fig:posts_mc}
\end{figure}


\subsection{Constraints on mass and concentration of the subhalo}\label{sec:mc_constraints}
    
To compare the inferred concentration of the subhalo to $\Lambda$CDM constraints, we define a derived parameter to be the halo concentration $c_{200} \equiv r_s/r_{200}$, where $r_{200}$ is the approximate virial radius the halo would have if it were in the field where tidal stripping is absent. Joint posteriors in $m_{200}$ and $c_{200}$ are plotted in figure \ref{fig:posts_mc}, where the red and blue shaded regions refer to model tNFW (the elliptical model) and tNFWmult (which has added multipoles) respectively. Note that in the tNFW model, a mass of $\sim 5\times10^9 M_\odot$ is preferred with a minimum concentration of several hundred, whereas in tNFWmult, a mass of $\sim 3\times 10^{10}M_\odot$ is preferred with minimum concentration of 60. For an initial comparison we plot the mass-concentration relation of field halos from \cite{dutton2014} (black solid line) along with the 2$\sigma$ scatter in concentration (black dashed lines). As the figure suggests, we find the inferred concentration in the tNFWmult model exceeds the median concentration for CDM field halos by at least $\approx 5\sigma$, where $\sigma$ is the posterior uncertainty in the concentration.

However, comparing to field halos at the lens redshift is insufficient, since it is well known that subhalos are denser than similar mass halos in the field, for two reasons: 1) subhalos fell in earlier when the critical density of the Universe was higher; 2) most subhalos have endured some tidal stripping, which preferentially removes more mass from the outer regions and renders the subhalo denser overall compared to unstripped halos of similar mass. 
To directly test against CDM predictions, we turn to cosmological simulations and examine the subhalo population in simulated galaxies analogous to the lensing galaxy in J0946+1006, to evaluate whether our subhalo is truly an outlier that cannot be explained with CDM. We will do so using IllustrisTNG in Section \ref{sec:illustris}.

\subsection{Interpretation in terms of the subhalo's perturbation radius and estimated robust mass}\label{sec:rpert}

It is evident from Figure \ref{fig:posts_mc} that the model used for the host galaxy can influence the inferred properties of the subhalo significantly: when the primary lens galaxy is allowed to have non-elliptical contours by adding multipoles, the subhalo prefers to be more massive but less concentrated. This begs the question, if the primary lens model is refined further---perhaps by a multi-component model or allowing for twist of isodensity contours---might the inferred subhalo have a sufficiently low concentration to be consistent with CDM?

To answer this question, it is essential to have an understanding of what is actually being constrained about the subhalo. In \cite{minor2017} it was shown that for perturbing subhalos, one can robustly infer the subhalo's projected mass enclosed within its \emph{perturbation radius} $r_{\delta c}$, defined as the distance from the subhalo center to the point on the critical curve that is being perturbed the most. Provided the detection is of high significance, this mass estimate is robust to changes in the subhalo's assumed density profile and tidal radius (unlike the total subhalo mass which can be quite sensitive to these systematics). While approximate formulas can be derived for the perturbation radius $r_{\delta c}$ under the assumption of an isothermal (or truncated isothermal) profile, in the case of an NFW profile $r_{\delta c}$ must be found numerically via a root finding algorithm. For example in model tNFW, the best-fit model recovers $r_{\delta c} \approx 0.36'' \approx 1.3$ kpc, whereas Model tNFWmult gives $r_{\delta c} \approx 0.31'' \approx 1.1$ kpc. The primary reason for the difference is that the position of the subhalo differs slightly in each model, the difference being comparable to a single pixel length ($0.05''$). In order to compare different models, we will evaluate the projected mass within 1 kpc as being robust to good approximation. The inferred robust mass for tNFW is $M_{2D}$(1kpc)$ = (2.5\pm0.3)\times 10^9M_\odot$, while model tNFWmult has $M_{2D}$(1kpc) $ = (3.3\pm0.3)\times 10^9M_\odot$, where the uncertainties give the 95\% credible intervals.\footnote{More precisely, the robust quantity is $M_{2D}(r_{\delta c})/\alpha$, where $\alpha$ is the slope of the host halo's density profile at the position of the subhalo (as discussed in \citealt{minor2017}). This implies that the inferred subhalo mass can be biased if the inferred $\alpha$ is biased. However, among our four lens models we fit, the inferred slope $\alpha$ are all consistent with each other within 90\% CL, suggesting the inferred slope is not sensitive to bias due the subhalo model being fit. If, however, the true density slope is shallower than our lens models are finding, the inferred subhalo mass would be reduced. For example, if the slope is close to isothermal ($\alpha \approx 1$), as found in \cite{collett2014}, this would reduce our subhalo masses by roughly 25\%. Modeling both the subhalo \emph{and} the second lensed source with a sufficiently flexible host galaxy model may shed light on whether the density slope is biased here or not.}

\begin{figure*}
	\centering

	\subfigure[elliptical models]
	{
		\includegraphics[height=0.4\hsize,width=0.45\hsize]{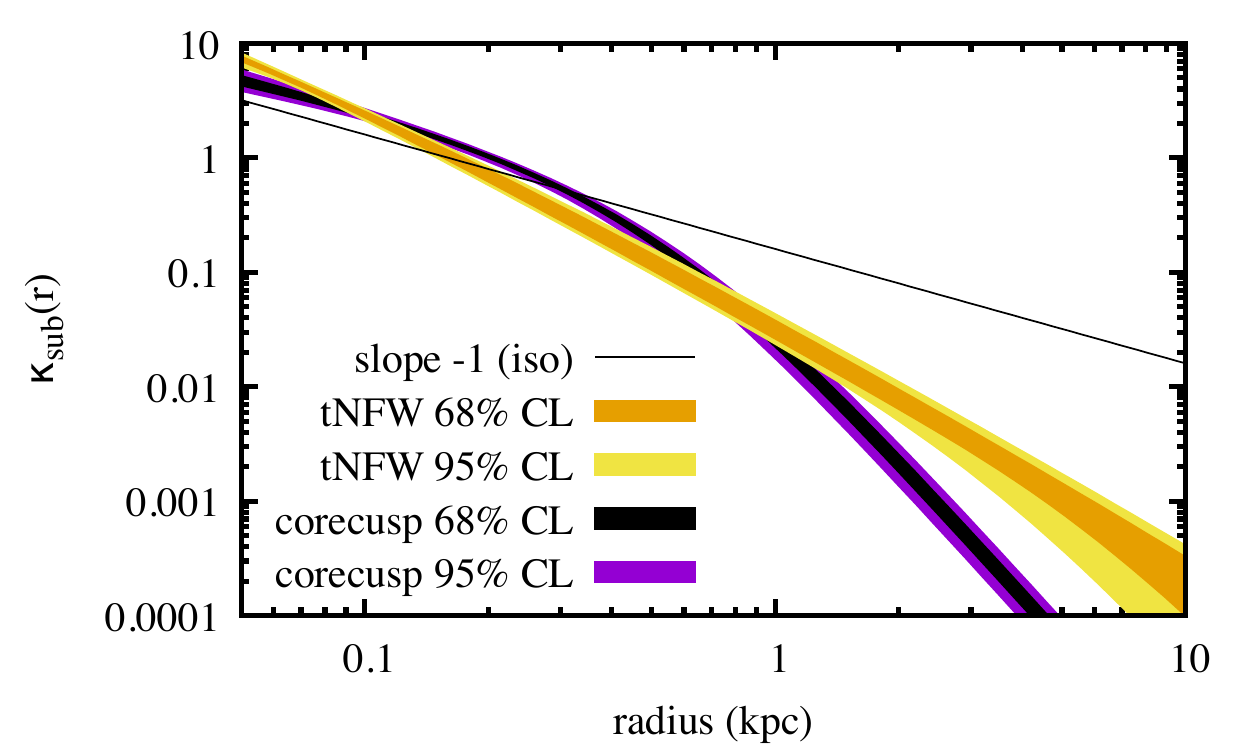}
		\label{fig:kprofile_tnfwcc}
	}
	\subfigure[Models w/ multipoles]
	{
		\includegraphics[height=0.4\hsize,width=0.45\hsize]{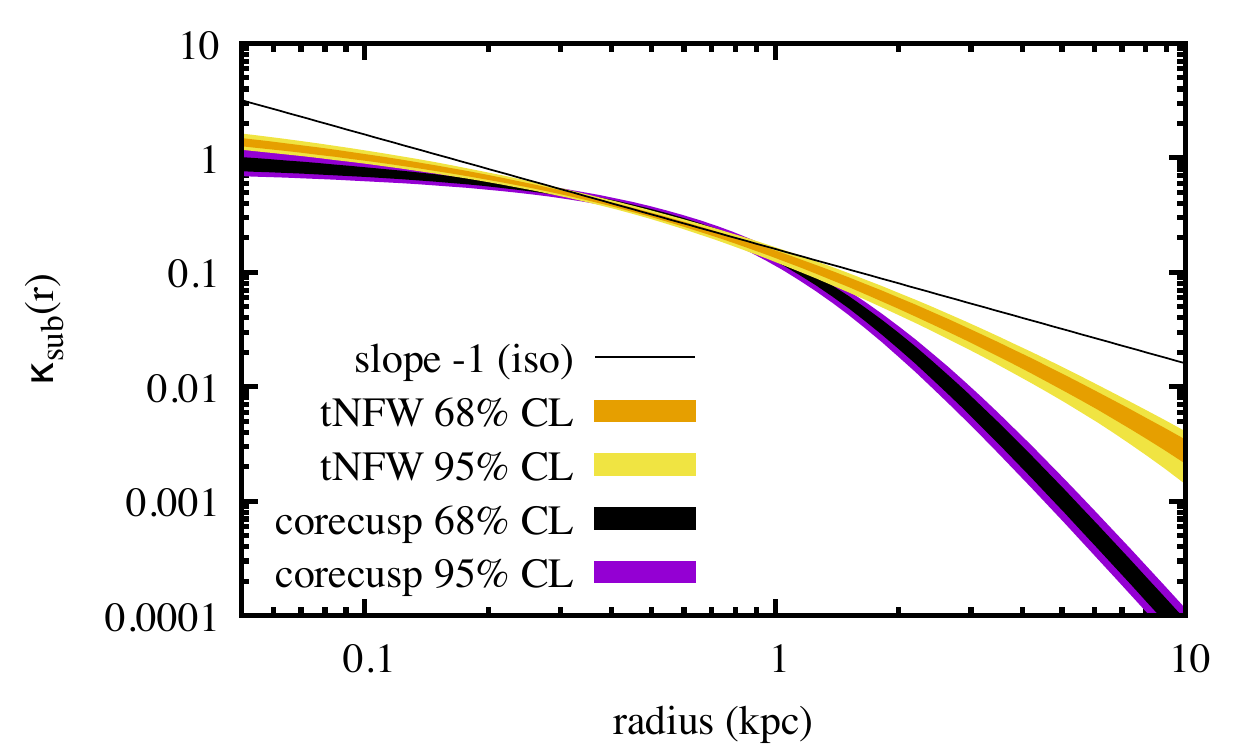}
		\label{fig:mprofile_tnfwccmult}
	}
	\caption{Projected density profiles, for two different subhalo models (tNFW and corecusp). Figure (a) shows the models with an elliptical primary lens, while figure (b) shows the models where multpoles are added to the primary lens galaxy to capture deviations from ellipticity. The inferred average log-slope from $r=0.3$ kpc to $r=1$ kpc is similar regardless of the subhalo profile used.}
\label{fig:profiles_tnfwcc}
\end{figure*}

To verify that the subhalo's mass enclosed within 1 kpc is being robustly inferred in our models, we start with our best-fit models and vary the subhalo mass $m_{200}$ over a table of values; for each $m_{200}$, we solve for the required concentration parameter $c_{200}$ to keep the perturbation radius $r_{\delta c}$ (and with it, the robust mass) constant. The corresponding curves are plotted for Models 1 and 2 in Figure \ref{fig:posts_mc} (red and blue curves respectively). Note that the posteriors follow essentially the same trend as the curves do, indicating that the mass within 1 kpc is being kept approximately constant over the parameter space covered by the posterior.

The inferred position of the subhalo, and hence its perturbation radius, is unlikely to differ by more than the PSF full width at half maximum, approximately two pixel lengths ($\sim 0.1''$). Thus it is reasonable to expect that any lensing solution with an equally strong fit will produce a posterior that lies close to the band swept out by the blue/red curves in Figure \ref{fig:posts_mc}. This implies that if a solution was found that allowed for a low enough concentration ($\lesssim 20$) to be consistent with CDM, the subhalo's $m_{200}$ would have to be well above $10^{11}M_\odot$. With such a high mass, the subhalo's stellar component would likely be visible. For example, from Figure 10 of \cite{behroozi2013} the median stellar mass in a galaxy with halo mass $2\times10^{11}M_\odot$ is estimated to be $M_* \sim 1.6\times10^9M_\odot$, reaching down to $M_* \sim1.2\times10^9M_\odot$ for the 16th percentile of galaxies. Since the subhalo is conservatively estimated to have a luminosity no greater than $L_V\approx1.2\times10^8L_\odot$ (see section \ref{sec:lv_constraint}), this would require a V-band stellar mass-to-light ratio greater than 10$M_\odot/L_\odot$, which would be extraordinarily high. The requirement becomes even more stringent under the assumption that the perturber is a halo along the line of sight to the lens, as we will demonstrate in Section \ref{sec:los}. We will return to these arguments in Section \ref{sec:illustris} in the context of a more rigorous comparison to CDM subhalos using the Illustris TNG100-1 simulation.

\begin{figure*}
	\centering

	\subfigure[projected density profile]
	{
		\includegraphics[height=0.4\hsize,width=0.45\hsize]{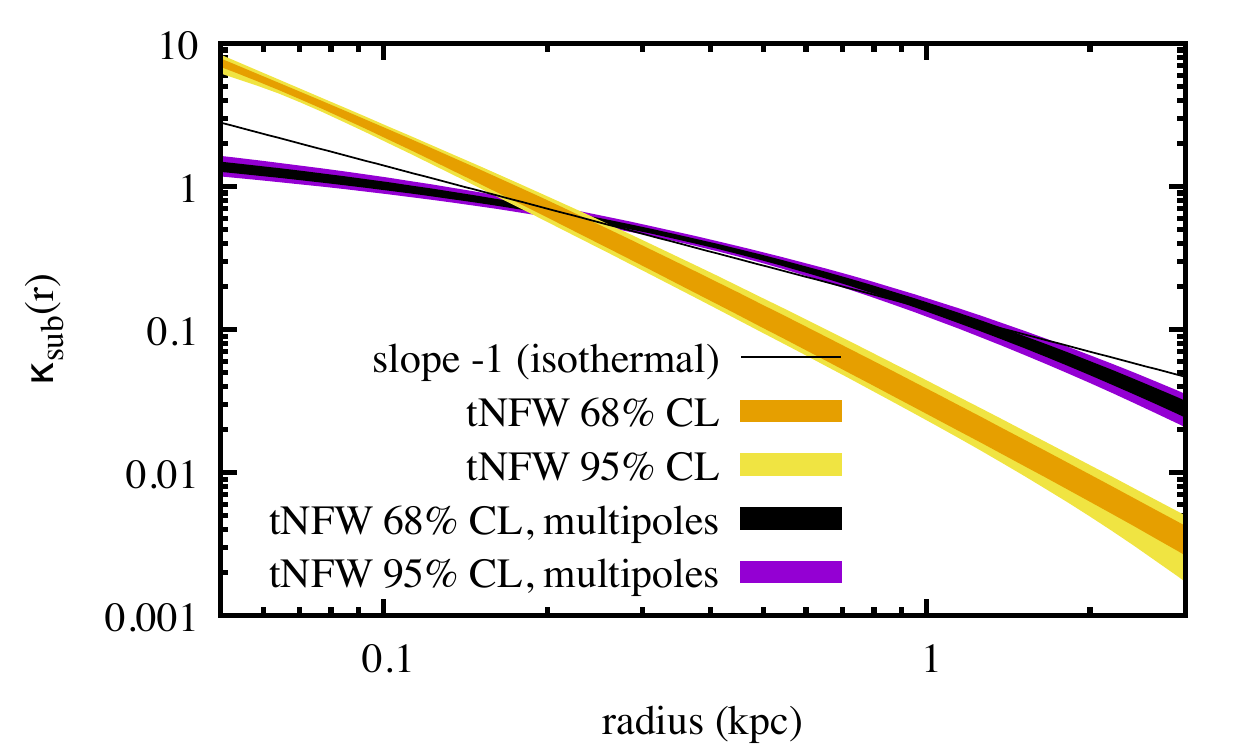}
		\label{fig:kprofile_tnfwmult}
	}
	\subfigure[mass profile]
	{
		\includegraphics[height=0.4\hsize,width=0.45\hsize]{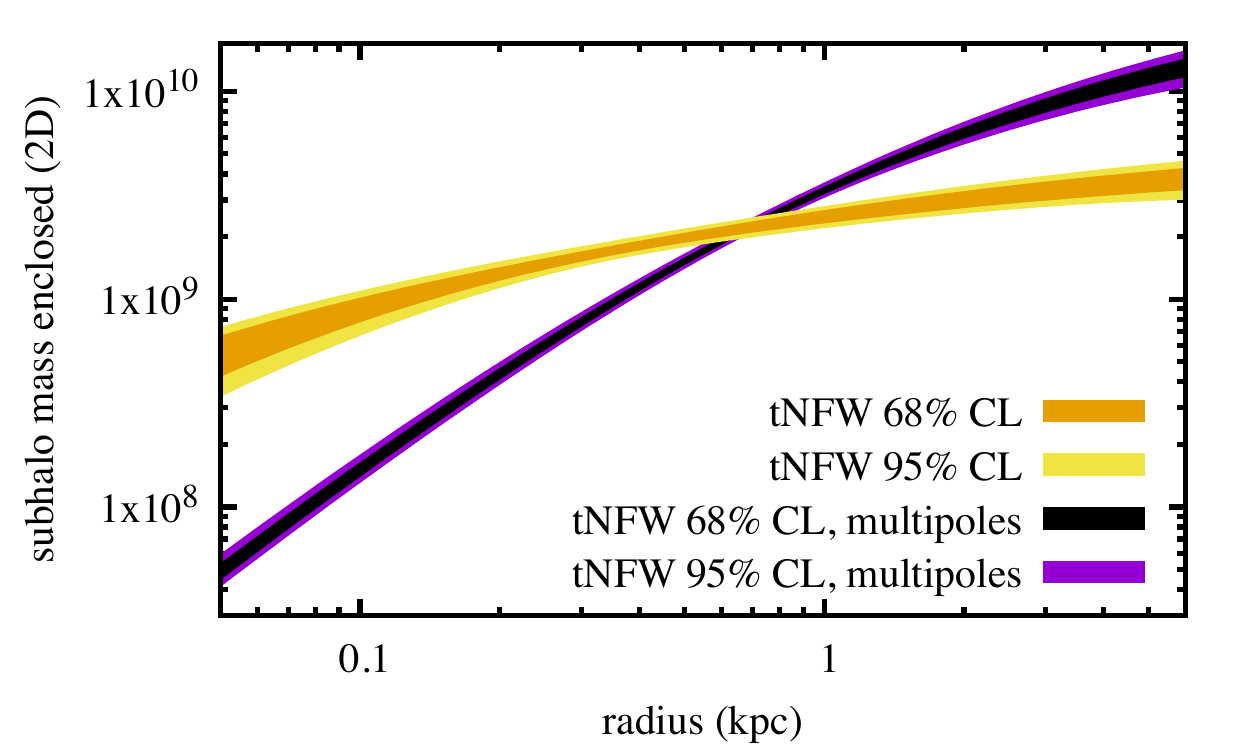}
		\label{fig:mprofile_tnfwmult}
	}
	\caption{Projected density and mass profiles, for truncated NFW models with versus without multipoles added to the primary lens galaxy's projected density. Note from figure (b) that the solutions have similar projected masses around 1 kpc, even though the profile shape and the total subhalo mass is rather different in either case.}
\label{fig:kmprofile}
\end{figure*}

\subsection{Characterizing the constraints on the subhalo's density profile}\label{sec:logslope}

To get a sense of how well the density profile is being constrained, we redo the analysis with a different density profile for the subhalo, this time the ``cuspy halo model'' of \cite{munoz2001} (this is also the ``corecusp'' model in \cite{andrade2019} with core size set to zero). In this model, the parameters are the inner and outer slopes and the break radius $a$. We fix the outer slope to -5 to be consistent with our tNFW runs, and vary the inner slope $\gamma$ and break radius $a$ freely. (Note that in the special case $\gamma = 2$, the profile is quite similar to the Pseudo-Jaffe profile used in \cite{vegetti2010} except with outer slope -5 instead of -4.) As with the tNFW fits, we run two versions: one with an elliptical host galaxy (simply labeled ``CoreCusp''), and one with multipoles (labeled ``CoreCuspmult''). The inferred parameter values and uncertainties are given in Table \ref{tab:models}. In Figure \ref{fig:profiles_tnfwcc} we plot the inferred projected density profiles by showing the 68\% and 95\% probability region of the projected density calculated over a large number of radial bins. Figure \ref{fig:profiles_tnfwcc}(a) shows the density profiles for the corecusp and tNFW models with elliptical host, while Figure \ref{fig:profiles_tnfwcc}(b) shows the corresponding profiles with multipoles added. Note that in either case, the inferred density and its log-slope within the interval between approximately 0.5 kpc and 1 kpc are approximately consistent with each other; in the case with multipoles, the log-slopes are both approximately isothermal over this interval, although they steepen as they reach 1 kpc.

For the purpose of comparing to the IllustrisTNG results in Section \ref{sec:illustris}, we define a derived parameter $\gamma_{2D,avg}$ to be the average log-slope from 0.75 kpc to 1.25 kpc. In the tNFW model, $\gamma_{2D,avg}$ is largely determined by the scale radius or concentration parameter. Hence, we see that, to first approximation, both $M_{2D}$(1kpc) and $\gamma_{2D,avg}$ are determined robustly regardless of the profile, which are useful physical quantities to test against subhalos in cosmological simulations. Note that this does \emph{not} mean that $\gamma_{2D,avg}$ is coming out exactly the same in tNFW vs tNFWmult. In Figure \ref{fig:kmprofile}(a) we plot the tNFW models with and without the multipoles for direct comparison. Note the slopes are significantly shallower in the models that include multipoles. This is directly related to the fact that tNFWmult allows for lower concentrations compared to tNFW. In Figure \ref{fig:kmprofile}(b) we plot the corresponding projected mass profiles; here it is clear that the masses are roughly similar near 1 kpc. Since the perturbation radius is approximately 1 kpc, this is consistent with the conclusion of \cite{minor2017} that the subhalo's projected mass within this radius is a robust measurement. (In this case the masses become equal around 0.7 kpc; this is because the position of the subhalo is not exactly the same in either case, differing by about a pixel length, which is discussed in Appendix \ref{sec:appendix_lensmodels}).

\subsection{Revisiting the upper bound on the subhalo's stellar luminosity}\label{sec:lv_constraint}

\begin{figure*}
	\centering
	\subfigure[F814W image]
	{
		\includegraphics[height=0.46\hsize,width=0.48\hsize]{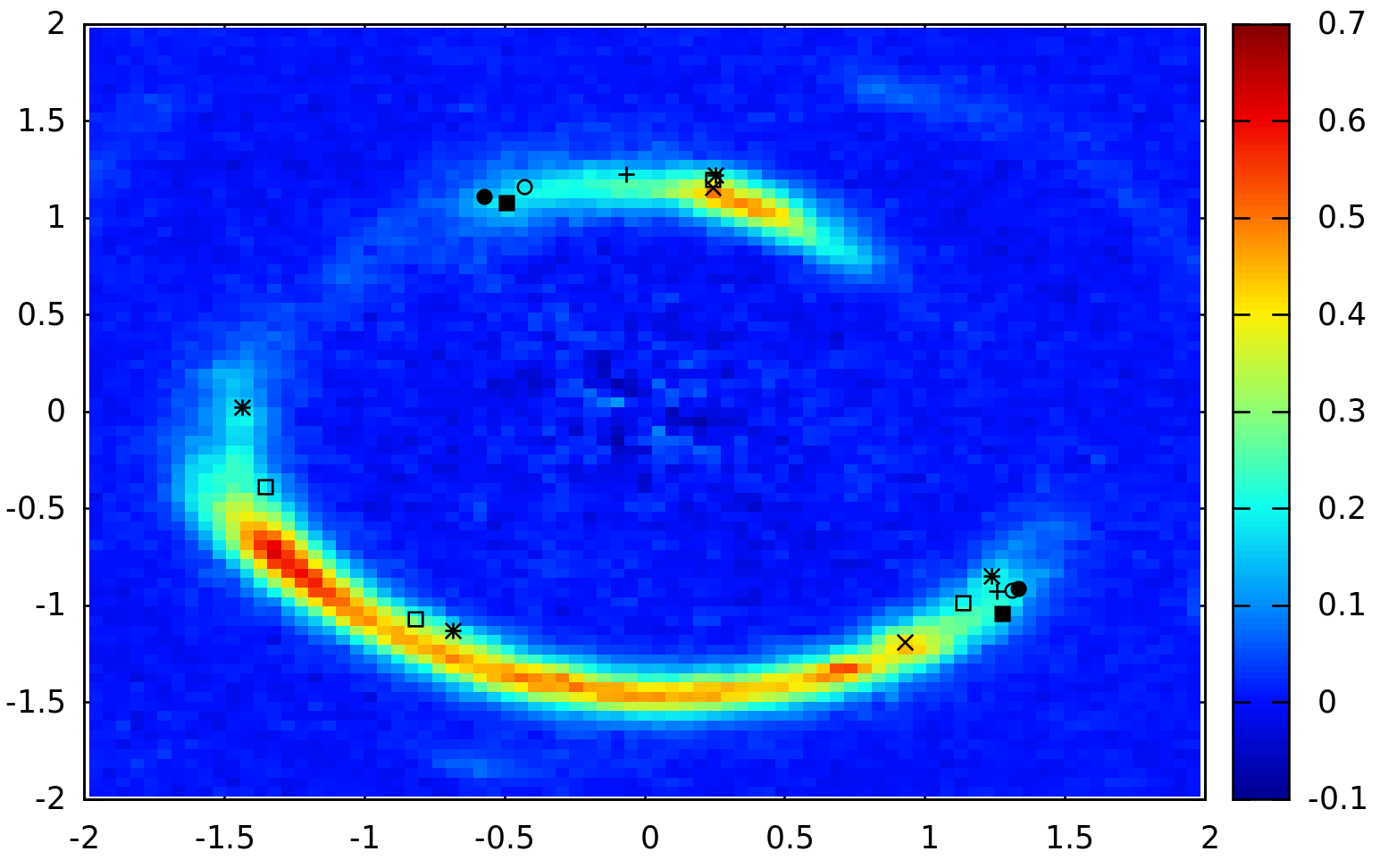}
		\label{f814w}
	}
	\subfigure[F336W image]
	{
		\includegraphics[height=0.46\hsize,width=0.48\hsize]{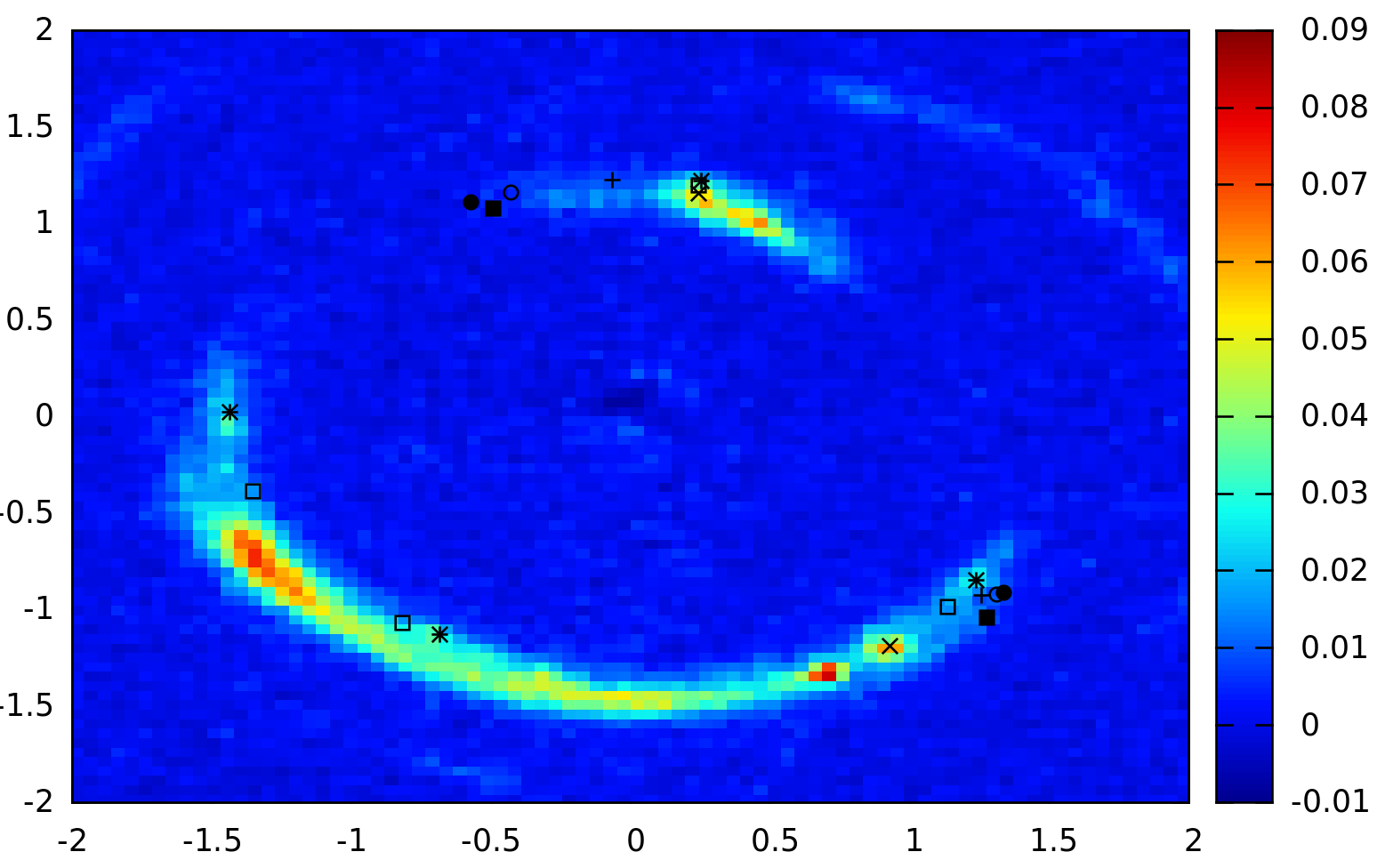}
		\label{f336w}
	}
	\caption{Image of SDSSJ0946+1006 in two different HST filters: F814W on the left (from \citealt{gavazzi2008}), which is used for the primary lensing fits in this paper; and F336W on the right, from \cite{sonnenfeld2012}. Both images have the foreground galaxy subtracted out. To illustrate the lens mapping near the subhalo, seven points in the top arc are shown (black points with various markers) along with their partner images on the bottom arc according to our best-fit lens model (tNFWmult). Note that in F336W, the surface brightness is dramatically reduced near the subhalo (filled circle, filled square, and open circle), but this is also true for the partner images in the lower right, indicating the lens mapping is consistent in both bands.} 
\label{fig:multiband}
\end{figure*}

In \cite{vegetti2010}, an upper limit on the subhalo's luminosity $L \lesssim 5\times10^6L_\odot$ is derived based on the residuals after subtracting their best-fit lens model. However, searching for a luminous signal in the residuals carries the risk that the lensed images from the model may have absorbed part of the subhalo's light and subtracted it out; this is especially true for pixellated sources which have a fair amount of freedom to absorb local fluctuations in surface brightness. In this section we revisit this bound and consider whether foreground light may be disguised by the lensed image near the subhalo.

One way to check whether the subhalo's foreground light may be present is to test the lens model in different bands. In \cite{sonnenfeld2012} this lens is modeled in several different HST bands (without including a substructure). For comparison, in Figure \ref{fig:multiband} we plot their image using the F336W filter with the foreground galaxy subtracted out, side-by-side with the foreground-subtracted F814W image from \cite{gavazzi2008} used in our lens modeling. In the shorter-wavelength F336W band there is a much smaller signal from foreground light, and indeed there are clearly fewer residuals from the foreground subtraction. In addition, the bright ``knob'' near the subhalo in the F814W image is barely visible in the F336W image. To test whether this feature may have a component of foreground light, we plot seven points in the vicinity of the subhalo (black points with varous markers) and shown their partner images on the bottom arc using our best-fit tNFWmult model. Note that although the surface brightness is dramatically reduced very close to the subhalo (filled circle, filled square, open circle), this is also true for the partner images in the lower right. This suggests that the variation in surface brightness is due to the source galaxy itself, rather than any disguised foreground light near the subhalo. This is a useful consistency check for our lens model, and suggests that if the subhalo's stellar light is visible in the image, it is clearly subdominant compared to the lensed source.

To place more rigorous constraints on the subhalo's luminosity, the most direct approach is to include a light profile for the subhalo in our model whose parameters are varied simultaneously with the lens/source parameters. We carry out this strategy by modeling the subhalo's light with a Gaussian profile, varying the total luminosity $L_V$ and half-light radius $r_{1/2}$ as free parameters. The surface brightness from the subhalo is added to the light from the lensed sources before convolving with the PSF. We allow the half-light radius to vary from 0.005'' up to 0.3'' (or from $\approx$ 18 pc up to $\approx$1.1 kpc). In all other respects, our model is identical to the tNFWmult model. 

Since the light profile approaches a point source at the low end of the prior range in $r_{1/2}$, numerical error can arise since the extent of the profile is small compared to the size of an image pixel. To solve this issue, we implement a ``zoom'' feature where any pixels whose borders lie close enough to the subhalo light (we choose the distance criterion to be within twice the half-light radius) are split into subpixels, such that the subpixels have lengths no longer than $\frac{1}{4}r_{1/2}$ in either direction; if a pixel already satisfies this criterion, no splitting occurs. The surface brightness is then evaluated at each subpixel, and these values are averaged to find the surface brightness of each parent pixel. With our choice of subpixel size, the surface brightness is quite accurate and does not significantly improve with further splittings. (Finally, we note that while we are using the zoom feature for a foreground light profile here, it can also be used for compact lensed sources as well.)

The best-fit half-light radius of the subhalo is $r_{1/2} = 0.74_{-0.10}^{+0.15}$ kpc, while the stellar luminosity constraint is surprisingly stringent, giving $L_V = (1.1 \pm 0.1)\times 10^8L_\odot$ (the uncertainties here give the 95\% credible interval). In addition, the subhalo's dark matter constraints are slightly different than in tNFWmult: the best-fit $M_{2D}$(1kpc) is $\approx 2\times 10^9M_\odot$, which is significantly smaller, while the scale radius has shrunk considerably to $\approx 0.1$ kpc. The question remains, however, can the apparently large subhalo luminosity be trusted?

Comparing the residuals to those of our fiducial models (Figure \ref{modelimg1}) show that the residuals below the subhalo are much reduced by the inclusion of the stellar light profile. However, as discussed in Section \ref{sec:results}, these residuals---which are located a few pixel-lengths away from the inferred subhalo center---are likely to be due to unsubtracted foreground light from the primary lens galaxy. This can be seen in Figure 2 of \cite{sonnenfeld2012}, where they subtracted the foreground galaxy using an elliptical Sersic model: the residuals in question (the F814W image is plotted in row 2 of that figure) form part of a continuum that go all the way to the central region where the foreground galaxy is brightest. (The foreground subtraction in \citealt{gavazzi2008}, used in this paper, employed a more sophisticated B-spline technique to remove the residuals, making this continuum less apparent.) As a result, the true luminosity of the subhalo is likely to be lower than our solution suggests. We thus regard $L_V < 1.2\times10^8L_\odot$ to be a conservative upper bound on the subhalo's stellar luminosity. The true luminosity is likely to be lower than this, but need not be as low as the bound derived in \cite{vegetti2010}.

Finally, we note that \cite{collett2020} have recently discovered a third lensed source at redshift $z=5.975$ using the MUSE spectrograph, one of whose images is very close to the location of the perturbing subhalo. However, the partner image (the larger of the two) from this source is observed to the lower right of where the large arcs are, and there is no discernible signal in that region in the F814 image. Therefore, we do not expect this high-redshift source to affect our upper bound on the subhalo luminosity derived here.

\section{Comparison to subhalo populations within analogue lensing galaxies in the Illustris TNG100-1 simulation}\label{sec:illustris}

In the preceding sections, we have identified three properties of the subhalo in J0946+1006 for which we have constraints in order to fit the lensing data well. Conservatively, the following bounds are required: the projected mass within 1 kpc, $M_{2D}($1kpc$) \gtrsim 2\times10^9 M_\odot$; the average log-slope of the 2D density profile from 0.75 kpc to 1.25 kpc $\gamma_{2D} < -1$; and the V-band luminosity $L_V \textless 1.2\times10^8L_\odot$. To address the question whether such a highly concentrated subhalo is consistent with CDM, we now explore whether subhalos that satisfy these properties arise in cosmological simulations.

To find analogues of the main lensing galaxy, we require a cosmological simulation that includes the effects of baryons on the mass distribution of galaxies. 
This is partly because the radiative cooling of baryons and subsequent adiabatic contraction (AC) of the dark matter halo results in a denser galaxy, with larger Einstein radius, compared to the same galaxy in a DMO simulation \citep{blumenthal1986,gnedin2004}. The baryonic physics, in particular adiabatic contraction and AGN feedback, also affects the radial distribution and mass function of the subhalos \citep{despali2016}, which is critical for the comparison we are making here. We choose IllustrisTNG for this purpose, which is a suite of both hydrodynamical and DMO N-body simulations. The hydrodynamical suite of simulations includes the effect of gas cooling, as well as a prescription for subgrid physics including star formation/evolution, black hole seeding and evolution, and stellar and AGN feedback, while the DMO runs are strictly N-body simulations based on the CDM model \citep{nelson2018illustristng,Pillepich_2017,Springel_2017,Nelson_2017,Naiman_2018,Marinacci_2018}.

\subsection{Sample selection}\label{sec:illustris_sample}

\begin{figure*}
	\centering
	\includegraphics[height=0.4\hsize,width=1.0\hsize]{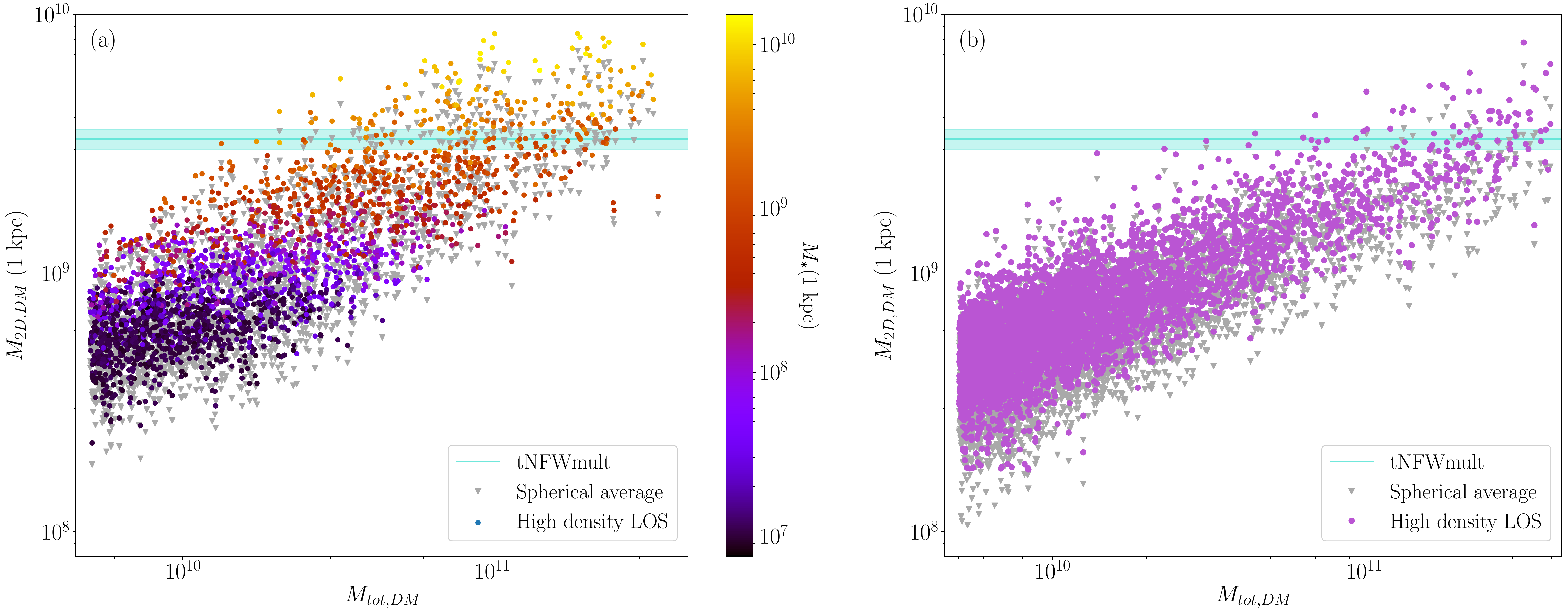}
	\caption{Projected mass within radius 1 kpc from subhalo center, plotted with respect to total mass for simulated subhalos in the hydrodynamical TNG100-1 run (left panel (a)) and the DMO TNG100-1-DARK run (right panel (b)). In both panels the gray triangle points are from the full spherical average over 1000 lines of sight (LOS), while the filled colored circles are the same subhalos averaged over the top 10\% of highest density in 1 kpc LOS. In panel (a), the circles are color-coded by $M_{*}$(1 kpc), while in panel (b) they are solid violet. The $2\sigma$ error bands for $M_{2D,DM}$(1 kpc) are also included for the ``tNFWmult'' lensing model, where the subhalo is modeled with a truncated NFW profile and our model allows for departures from ellipticity in the host galaxy. All the plotted subhalos lie within $r_{vir}$ from the host galaxy center. }
\label{fig:TNG_m2d_mtot}
\end{figure*}

Although our primary results will be obtained using a full hydrodynamical  simulation in IllustrisTNG, we will also analyze and compare to its DMO counterpart to gauge the effects of baryons on the satellite populations. We choose in particular the TNG100-1 and TNG100-1-DARK simulations, both of which have a box length of about 100 Mpc and the smallest dark matter particle mass of all the publicly available simulations in the TNG suite, with $m_{DM} = 7.5\times10^6M_\odot$ for the hydrodynamical run, and $m_{DM} = 8.9\times10^6M_\odot$ for the DMO run (TNG50 will have smaller mass particles but is not yet publicly available; \citealt{Nelson_2019}). Note that with this resolution, subhalos with masses less than $7.5\times 10^9M_\odot$ will have fewer than 1000 particles, and hence might be vulnerable to artificial disruption due to being unresolved \citep{carlsten2020radial}. For this reason, we restrict our search to subhalos with total dark matter masses greater than $5\times10^9M_\odot$ so as to exclude subhalos with less than $\sim$700 particles, to ensure all subhalos have $\sim$1000 particles or more. We do the same for the DMO run despite the slightly lower resolution (as we will see, however, among lower mass subhalos only a tiny fraction are expected to produce a large enough $M_{2D}$(1kpc), eliminating the subhalos with lower particle counts).

 We find analogue lensing galaxies in a manner similar to \cite{despali2016}, by selecting TNG100-1 hosts and subhalos at a redshift z = 0.2 with masses close to those measured for J0946+1006 \citep{auger2010}. We accomplish this by searching the TNG100-1 API for host halos with total masses $1\times10^{13}-6\times10^{13} M_\odot$ and stellar masses between $8\times10^{10}-10^{12} M_\odot$ to match the host (for the DMO run we only use the total mass criteria). From the hosts that match our search criteria, we select subhalos with total masses $8\times10^{9}-4\times10^{11} M_\odot$, and dark matter masses $5\times10^9M_\odot - 4\times10^{11}M_\odot$ to address the resolution issues aforementioned. In the DMO run, we only use the latter range. 

 Although our expectation that a strong candidate would have undergone significant tidal stripping suggests restricting our selection of subhalos to those closest to their host halos, we cannot predict where in the subhalo's orbit it is or when it passed pericenter as we restrict our analysis to one snapshot (or redshift). We therefore select all subhalos within $r_{vir}$ of their hosts, as they may have been stripped at some earlier redshift when at pericenter of their orbits. Additionally, we expect the subhalo to be close to its host in projection. This can be true for any subhalo depending upon the line of sight, further motivating relaxing the condition of distance from host. 

 This yielded a total of 3056 subhalos from 167 different hosts for the hydrodynamical run, and 4909 subhalos from 188 hosts in the DMO run. For the hydrodynamical run, 
 we further discarded subhalos that have less than 30 star particles, none of which had a projected mass within 1 kpc of at least $10^9M_\odot$; likewise, we discarded subhalos with less than $8.5\times10^8 M_\odot$ projected mass in 1 kpc, which is below the minimum requirement of $10^9 M_\odot$. This leaves us with a final sample of 2205 subhalos for the hydrodynamical run.
 
 For each subhalo, a spherically averaged density profile is constructed by binning the dark matter particles in radial mass bins, and 2D mass and surface density profiles are constructed similarly by binning in cylindrical shells and averaging over 1000 random lines of sight (LOS). Previous studies have pointed out that due to the triaxiality of dark matter halos, lines of sight along the long axis of a halo can result in a significantly higher projected density and thus a stronger lensing signal due to selection bias (\cite{Sereno_2015}). In view of this, the strongest subhalo perturbations may be generated by subhalos whose long axes are oriented nearly along our line of sight. In the following analysis, we explore this possibility by selecting the top 10\% of LOS which return the highest projected density within 1 kpc and produce profiles by averaging over these LOS for each subhalo, and contrasting their properties to the sample averaged over all LOS.
 
 Finally, to obtain $\gamma_{2D}$ and circumvent the issue of noise near the inner regions of the profiles of these subhalos, we fit the profile with a power law from $0.75 - 1.25$ kpc, and determine the slope for both the full spherical average, as well as the high density LOS. It should be borne in mind, however, that we are probing a regime in which the resolution is not sufficient to produce smooth density profiles for many subhalos; although our averaging procedure mitigates the noise, this may still lead to some artificial scatter in the inferred subhalo slopes. \footnote{The Illustris TNG50 simulation will solve this issue, at the cost of a smaller sample of subhalos due to smaller box size. The higher resolution of TNG50 will be able to better resolve the stellar profiles, as well as better resolution in mass by an order of magnitude and improved spatial resolution by a factor of $\sim$2 compared to TNG100-1 (\cite{Pillepich_2019}). A reanalysis combining the results here of TNG100-1 with those from TNG50 would be useful for more robust measurements of the inner slope of the density profile, while TNG100-1 gives us access to a larger population of subhalos in this mass range (\cite{Nelson_2019}).}
 
 \begin{figure*}
	\centering
	\includegraphics[height=0.4\hsize,width=1.0\hsize]{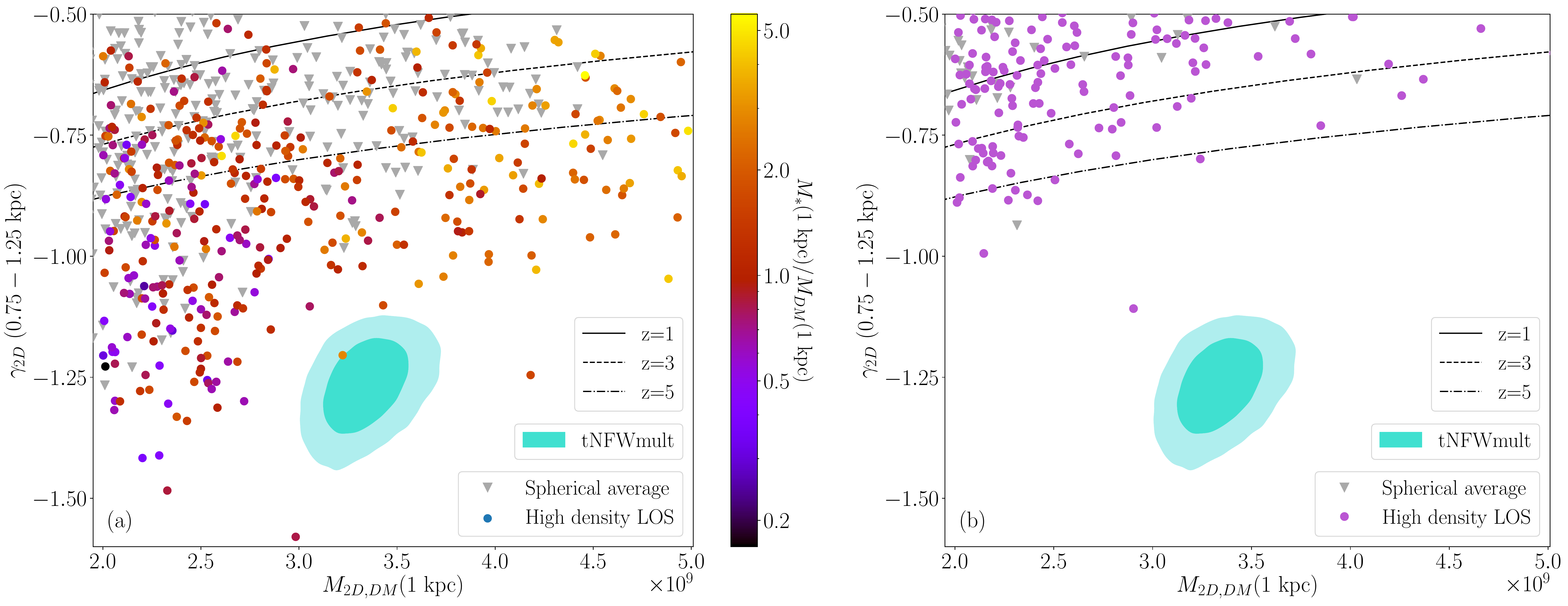}
	\caption{Slope of the surface density profile versus projected mass in 1 kpc for the TNG subhalos in the range $M(1$ $\mathrm{kpc}) = 2-5\times10^9 M_\odot$ and $\gamma_{2\mathrm{D}}\leq-0.5$ for both the hydrodynamical (panel (a)) and DMO (panel (b)) runs. Plotted are the full spherical average over 1000 lines of sight (gray triangles) as well as the average over the top 10\% highest density LOS in 1 kpc, color-coded by $M_*(1 \mathrm{kpc})/M_{DM}(1 \mathrm{kpc})$ in panel (a), and solid violet for panel (b). Both panels include the 68\% and 95\% contours from lensing constraints for our most conservative model, in which the subhalo is modeled with a truncated NFW profile and our model allows for departures from ellipticity in the host galaxy (which we term the ``tNFWmult'' model). and the cosmological relation for dark matter for z = 1, 3, 5 at $3\sigma$.}
\label{fig:TNG_subs_contours_cosmoRels}
\end{figure*}

\subsection{Subhalo candidates and comparison to lensing constraints}\label{sec:illustris_results}

 In Figure \ref{fig:TNG_m2d_mtot}, we plot the projected DM mass of each subhalo within 1 kpc, $M_{2D,DM}$(1kpc), versus the total subhalo mass for both the hydrodynamical (panel (a)) and DMO (panel (b)) simulation runs. The gray triangle points represent the full spherical average over 1000 random LOS for each subhalo, while the circles color-coded by $M_{*}$(1kpc) represent the average over the top 10\% of the highest density LOS for each subhalo. Note that the required $M_{2D,DM}$(1kpc) is achieved only for the subhalos of total mass $\sim10^{10}M_\odot$ or greater. The remaining subhalos with $M_{2D,DM} (1 \mathrm{kpc}) < 10^9 M_\odot$ are eliminated from the proceeding analysis. This cut also eliminates most of the subhalos with low stellar content below $10^8 M_\odot$ within 1 kpc. The color coding for the circle points reveals a trend in which the stellar mass in 1 kpc increases for larger projected halo mass within 1 kpc. Consistency with the lensing results (given by the horizontal shaded bar) therefore requires at least $\sim 2\times 10^9M_\odot$ of stellar mass within 1 kpc, despite the observed subhalo luminosity implying a much smaller stellar mass, a point to which we will return in section \ref{sec:discussion}.


In panel (b) of Figure \ref{fig:TNG_m2d_mtot}, we plot all 4909 subhalos found in TNG100-1-DARK. The trend of the points is similar to panel (a), however with a shallower slope. Additionally, less of the subhalo population falls in the $2\sigma$ band for $M_{2D}$(1kpc) than the hydro run in panel (a), and those that fall in, do as a result of selection bias, which is visible in a comparison between the violet circle points and the full spherically averaged gray triangle points: the high density averaged ones (violet) obtain a large boost in projected mass within 1 kpc compared to the gray triangles. The large boost implies selection bias is amplified in DMO subhalos, which are more spherical in hydrodynamical simulations as a result of the added central potential from the baryons \citep{Dubinski_1994,kazantzidis2004}. Additionally, the trend is less steep in panel (b) than in panel (a), indicating subhalos are denser in the hydrodynamic run, particularly for higher subhalo masses.
 
 Next, we examine the density slope constraints. In Section \ref{sec:logslope}, we showed that the log-slope of the subhalo's projected density profile is well constrained from the lensing signal over the approximate range from 0.1 up to 1 kpc. However, since the surface density profiles of the subhalos in TNG100-1 are often noisy near the central regions within 1 kpc due to poor resolution (see Appendix), we pick the upper end of this range and fit a line to the binned profiles between 0.75 and 1.25 kpc to obtain the average log-slope $\gamma_{2D}$ for each subhalo.
 
 In Figure \ref{fig:TNG_subs_contours_cosmoRels} we plot the average log-slope of the projected density profiles, $\gamma_{2D}$, with respect to $M_{2D}$(1kpc), for all subhalos with $2\times 10^9 < M_{2D,DM}(1\mathrm{kpc}) < 5\times 10^9 M_\odot$ and $\gamma_{2D} < -0.5$. Here again, panel (a) corresponds to the hydrodynamical run, while panel (b) is the DMO run, and the gray triangles again represent the full spherical average over 1000 LOS, whereas the colored circles are the average over the top 10\% of the highest density LOS for each subhalo. In Figure \ref{fig:TNG_subs_contours_cosmoRels}(a) the circles are color coded by the ratio of stellar mass over dark matter mass within 1 kpc of the subhalo's center. As in Figure \ref{fig:TNG_m2d_mtot}, the line of sight is a significant factor, with the gray triangle markers being significantly shifted towards shallower slopes and smaller $M_{2D,DM}$(1kpc) from the color-coded points. This effect is more significant for the DMO run, which is again consistent with a greater triaxiality expected for the DMO halos \citep{Dubinski_1994, kazantzidis2004}. The corresponding lensing constraints for the slope and the projected mass within 1 kpc are also plotted as contours at $68\%$ and $95\%$ CL. We show only our most conservative model ``tNFWmult'', which modeled the subhalo with a truncated NFW profile and had the highest Bayesian evidence among the lens models we fit. A similar comparison that includes the other lens models, which have even steeper slopes compared to the model shown, is shown in Appendix \ref{sec:appendix_lensmodels}.

 We also include the cosmological relation for dark-matter-only field halos with concentrations 3$\sigma$ above the median for redshifts $z=1,3,5$ from \cite{dutton2014}. Note that even at the $3\sigma$ level, typical CDM field halos with masses within this range generally produce slopes shallower than -1; this can be seen in Figure \ref{fig:TNG_subs_contours_cosmoRels}(b), where aside from some scatter, subhalos generally reside in the region where slopes are shallower than -1, in contrast with Figure \ref{fig:TNG_subs_contours_cosmoRels}(a) which shows the hydrodynamical subhalo population having steeper slopes. This suggests that for many of the subhalos that achieve $M_{2D}$(1kpc) in the hydrodynamical run, tidal stripping and adiabatic contraction has steepened the profiles by contracting more dark matter into 1 kpc. However, in the vast majority of cases this does not result in a steep enough slope to fit the lensing constraints. Roughly $\sim60$ of the hydrodynamical subhalos are steep enough to be consistent with our inferred slope constraints when averaged over the top 10\% of high density LOS (colored circles), while only 6 subhalos are steep enough when measured from the full spherical average (gray triangles). In addition, none of the subhalos are consistent with both the inferred slope and $M_{2D}$(1kpc) when averaged over all LOS. The difference is starker in the DMO simulation (panel (b)), where none of the subhalos produce a steep enough slope even for the highest density LOS.

 \begin{figure}
	\centering
	\includegraphics[height=0.8\hsize,width=1.0\hsize]{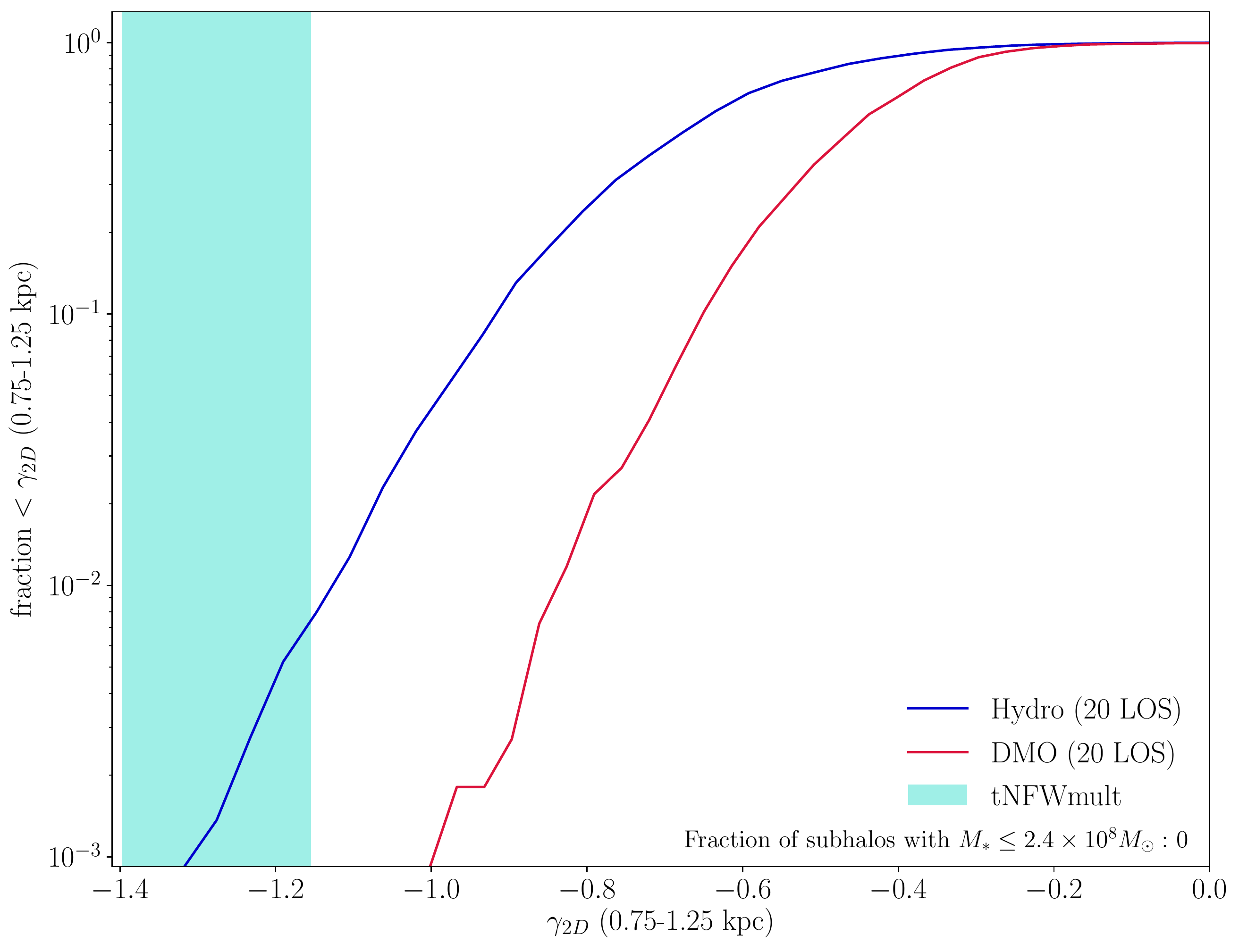}
	\caption{Fraction of mock ``observations'' in our TNG100 sample for which a subhalo's projected density log-slope is steeper than a given threshold $\gamma_{2D}$(0.75-1.25kpc). Here we include only the observations that produce a projected mass within 1 kpc $M_{2D}$(1kpc) in the range $3-4\times10^9M_\odot$, for comparison to our fiducial lensing model (``tNFWmult''); the 95\% credible interval in $\gamma_{2D}$ from our lens model is shown as the vertical shaded bar. (Joint constraints in $M_{2D}$(1kpc) and $\gamma_{2D}$ are shown in Figure \ref{fig:TNG_subs_contours_cosmoRels}). For each subhalo, 1000 lines of sight were generated, and each mock ``observation'' consists of 20 similar LOS over which the profiles are averaged to reduce noise. Note than less than 1\% of the mock observations involve slopes that satisfy the lensing constraint. The fraction is even smaller for our other lens models (shown in Appendix \ref{sec:appendix_lensmodels}.)}
\label{fig:cumdist_slope_34}
\end{figure}

\subsection{Likelihood of a perturbing subhalo being as concentrated as the subhalo in SDSSJ0946+1006}\label{sec:illustris_slope_pct}

 Since the projected slope and mass can depend significantly on the line of sight, we can ask the following question: supposing one makes a large number of mock ``observations'' over many lines of sight for each subhalo in our TNG sample, what fraction of such observations would involve $M_{2D}$(1kpc) and $\gamma_{2D}$(0.75-1.25kpc) values consistent with our lensing constraint? This gives an idea of the likelihood of seeing a lensing perturbation by a subhalo as concentrated as the one in SDSSJ0946+1006. To address this question, it would be a mistake to use all 1000 individual lines of sight: the noisiness of the profiles affects the slope calculation, as discussed in Section \ref{sec:illustris_sample}. However, we find that averaging the profiles over as few as 20 nearby LOS reduces the noise dramatically (see Appendix \ref{sec:appendix_lensmodels} for comparison of different LOS averaging). Thus, we generate mock ``observations'' by averaging 20 nearby lines of sight. The procedure is carried out as follows: for each subhalo, we calculate $M_{2D}$(1kpc); any lines of sight whose $M_{2D}$(1kpc) lies outside the range $3-4\times10^9M_\odot$ favored by our fiducial lens model is discarded. The remaining lines of sight are sorted by their $M_{2D}$ values and divided into groups of 20 LOS (with a few groups possibly obtaining one extra LOS); each group constitutes a mock ``observation'', over which the profiles are averaged and the resulting slope $\gamma_{2D}$(0.75-1.25kpc) is calculated. This process is repeated for each subhalo in our TNG100-1 sample, both for the hydrodynamical and DMO simulations. Note that in referring to each ``observation'', we are not simulating the lensing perturbation of the subhalos here, but rather just calculating the projected mass and (smoothed) density slope of each simulated subhalo.
 
 In Figure \ref{fig:cumdist_slope_34} we plot the fraction of mock observations for which the projected (smoothed) density slopes are steeper than a given threshold $\gamma_{2D}$. The blue curve corresponds to subhalos in the TNG100-1 hydrodynamical simulation, whereas the red curve gives the DMO simulation. The vertical shaded bar gives the 95\% credible interval from the lensing constraint for our fiducial model (defined by the 2.5\% and 97.5\% percentiles of the marginalized posterior in $\gamma_{2D}$). Note that less than 1\% of the observations have projected density slopes consistent with the lensing constraint; in the DMO simulation, \emph{none} of the observations produce steep enough density slopes. The situation is even starker for our other lens models, for which not a single mock observation produced a steep enough slope to match the lensing constraints even in the hydrodynamical simulation (see Appendix \ref{sec:appendix_lensmodels} for the corresponding plots with all lens models).
 
 This result can be interpreted as follows. The angular size of a lensing perturbation is determined primarily by the subhalo's projected mass and proximity to the lens's critical curve \citep{minor2017}, whereas to first approximation, the density slope affects the perturbation's shape, but not its size (Minor et al., in prep). This means that any of the mock observations in Figure \ref{fig:cumdist_slope_34} would produce a lensing perturbation of similar angular size if the simulated subhalo were placed at the position of the subhalo in SDSSJ0946, and thus the majority of them are likely to be detectable. However, fewer than 1\% of these observations would produce a projected density slope as steep as what the lensing constraints are telling us. This suggests that the subhalo in SDSSJ0946+1006 is truly an outlier in CDM (at the $99\%$ confidence level) in terms of its central density slope, and hence concentration, given the size of the observed lensing perturbation. 

 Note that this argument can be made independently of the stellar mass of the subhalo. However, as we discuss in the following section, even those subhalos that do achieve densities and slopes that match our lensing constraints have stellar masses that far exceed the conservative upper bound on the subhalo's stellar mass, significantly worsening the tension with CDM.
 
\subsection{Discussion}\label{sec:discussion}

 Although the subhalo in SDSSJ0946+1006 is clearly an outlier in our subhalo sample in TNG100-1, as Figures \ref{fig:TNG_subs_contours_cosmoRels} and \ref{fig:cumdist_slope_34} show, we do find a few subhalos whose projected density profiles are consistent with the lensing constraints within 99\% CL for specific lines of sight. How are such high densities and density slopes achieved?
 
 To understand this, we first consider the simulated subhalos' stellar mass. Among the sample of subhalos with $M_{2D,DM}$(1 kpc) $\geq 2\times10^9 M_\odot$, Figure \ref{fig:TNG_m2d_mtot} shows that nearly all of them have total stellar masses greater than $10^9M_\odot$. Given the stellar luminosity constraint $L_V < 1.2\times10^8L_\odot$ found in Section \ref{sec:lv_constraint}, achieving a stellar mass this high would require the actual subhalo to have a V-band mass-to-light ratio greater than 8$M_\odot/L_\odot$, which would be extraordinarily high and would imply a very steep stellar initial mass function (IMF). In old stellar populations typical for dwarf satellites, a Salpeter-like IMF generally produces $M_*/L_V \approx 2M_\odot/L_\odot$ \citep{martin2008}. Direct measurements of the stellar IMF in Milky Way satellite galaxies infer a stellar IMF that is shallower than Salpeter, and hence imply even lower $M_*/L_V$ \citep{geha2013,gennaro2018}; this is also consistent with the expectation from scaling relations in local group dwarfs \citep{woo2008} and the low measured $M_*/L_V$ in low surface brightness dwarfs \citep{du2020}. While super-Salpeter IMF's have been deduced in high-redshift giant elliptical galaxies via spectroscopy and/or lensing \citep{conroy2012,newman2013}, an IMF extreme enough to produce $M_*/L_V > 8M_\odot/L_\odot$ has only been observed in the centers of massive galaxies \citep{conroy2017}, likely as a result of extreme starburst conditions \citep{chabrier2014} unlikely to occur in a dwarf galaxy. If we assume a value that is likely a typical upper limit for dwarf galaxies, $M_*/L_V \approx 2M_\odot/L_\odot$, our lensing constraint would imply $M_* \lesssim 2.4\times10^8M_\odot$.

Among the simulated subhalos in our sample that achieve $M_{2D}$(1kpc)$ \geq 2\times10^9M_\odot$ along some line of sight, \emph{none} have stellar masses equal to or smaller than our conservative upper bound of $2.4\times10^9M_\odot$. Since TNG100-1 has been shown to produce stellar masses roughly consistent with observations of dwarf galaxies in the Local Volume down to at least $M_{halo} \sim 10^{11}M_\odot$ \citep{carlsten2020luminosity}, this suggests that subhalos with such high projected masses must necessarily be accompanied by a stellar mass that exceeds that of the subhalo in SDSSJ0946+1006. However, the strong upper bound on the subhalo's stellar mass does not by itself disqualify the simulated subhalos in our sample, since we do not necessarily expect the simulation to reproduce the correct stellar mass within 1 kpc.  The stellar softening length $\approx 0.7$ kpc implies the stellar density within 1 kpc must be taken with a grain of salt \citep{Pillepich_2017}.
Nevertheless, all the subhalo candidates with steep enough slopes (the $<1\%$ of observations in Figure \ref{fig:cumdist_slope_34}) have at least $3\times 10^9 M_\odot$ in stellar mass within 1 kpc, an order of magnitude higher than our upper bound on the observed subhalo. Most also contain more stellar mass than dark matter mass within 1 kpc. This strongly suggests that the dark matter mass within 1 kpc has been boosted significantly by adiabatic contraction, an effect which may also be steepening the slope within 1 kpc \citep{gnedin2004}. This is supported by the fact that none of the mock ``observations'' produced a steep enough density slope in the DMO simulation (Figure \ref{fig:cumdist_slope_34}). Indeed, previous work suggests that IllustrisTNG has over-contracted dark matter halos at higher galaxy masses, producing high central dark matter fractions compared to observational constraints \citep{lovell2018}. (See Appendix \ref{sec:appendix_candidates} for a direct comparison of the stellar and dark matter density profiles for 21 of the closest candidate subhalos.) If the high concentration of these dark matter subhalos was achieved primarily through AC, then they are no longer consistent with the lensing results since the true stellar mass is likely to be below $3\times10^8M_\odot$, not enough for AC to be a significant factor in reality---again, unless the stellar mass-to-light ratio takes an unprecedented value $M_*/L_V \gtrsim 8M_\odot/L_\odot$.


We conclude that the subhalo in the lens SDSSJ0946+1006 appears to be a remarkable outlier in $\Lambda$CDM, in two respects. Its projected mass within 1 kpc, constrained by the lensing, is an outlier given the upper bound on its luminosity, unless its stellar mass-to-light ratio takes an extraordinarily high value (making it an outlier in stellar mass). In addition, its density slope is an outlier (at $>99\%$ confidence, by Figure \ref{fig:cumdist_slope_34}), unless a lens model could be found that produces a strong fit and infers a shallower density slope; but even if such a lensing solution could be found, this would imply a large halo mass and hence would again require a very high stellar mass-to-light ratio.

Perhaps the most plausible scenario for reconciling the subhalo with $\Lambda$CDM is if a good-fitting lensing solution could be found that produces a projected mass within 1 kpc $\lesssim 2\times 10^9M_\odot$. As Figure \ref{fig:TNG_m2d_mtot} shows, subhalos with $M_{2D} \approx 1.5\times10^9 M_\odot$ can have stellar masses reaching down to $10^8M_\odot$, which would make it consistent with our upper bound on the stellar mass. Since $M_{2D}$(1kpc) is robust to changes in the subhalo profile, however, this would likely require some change in the host galaxy model that would allow the subhalo to be placed at a slightly different position with respect to the critical curve. However, the trend we observe is that solutions with smaller $M_{2D}$(1kpc) seem to require a steeper slope to fit the lensing data, presumably to produce a strong enough perturbation. For example, our truncated NFW model (tNFW) model with a perfectly elliptical host galaxy finds $M_{2D}$(1kpc)$ \approx 2.5\times10^9M_\odot$, but requires a significantly steeper slope $\gamma_{2D} \approx -2$ (see Figure \ref{fig:TNG_subs_contours_cosmo_corecusp} in Appendix B, purple contour). As we shall see in Section \ref{sec:los}, the same problem arises if the perturber is assumed to be a line-of-sight halo, rather than a subhalo. Thus it remains to be seen whether an equally good solution with $M_{2D}$(1kpc) $\lesssim 2\times10^9M_\odot$ could be found that would allow a slope shallow enough to be consistent with the subhalos in Illustris TNG. Fitting both of the lensed sources (and perhaps even a third source recently detected by \citealt{collett2020}) with a flexible host galaxy and subhalo model, perhaps in combination with spatially resolved kinematics, may shed light on this by better constraining the host galaxy's density slope and contour shape in the vicinity of the subhalo.

\section{What if the perturber is a line-of-sight halo rather than a subhalo?}\label{sec:los}

\begin{figure*}
	\centering

	\subfigure[perturber mass within 1 kpc]
	{
		\includegraphics[height=0.38\hsize,width=0.48\hsize]{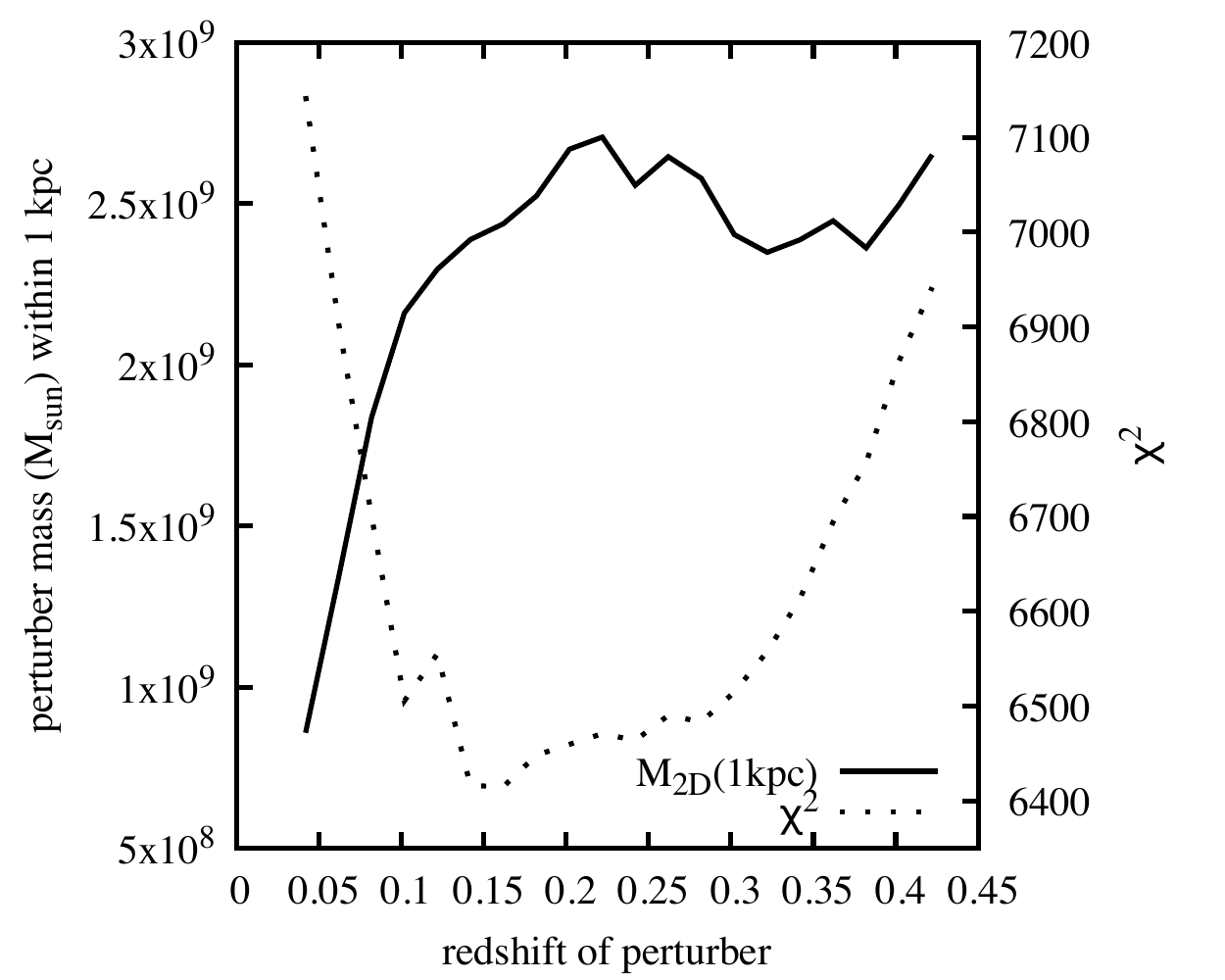}
		\label{fig:m1kpc_z}
	}
	\subfigure[perturber log-slope]
	{
		\includegraphics[height=0.38\hsize,width=0.48\hsize]{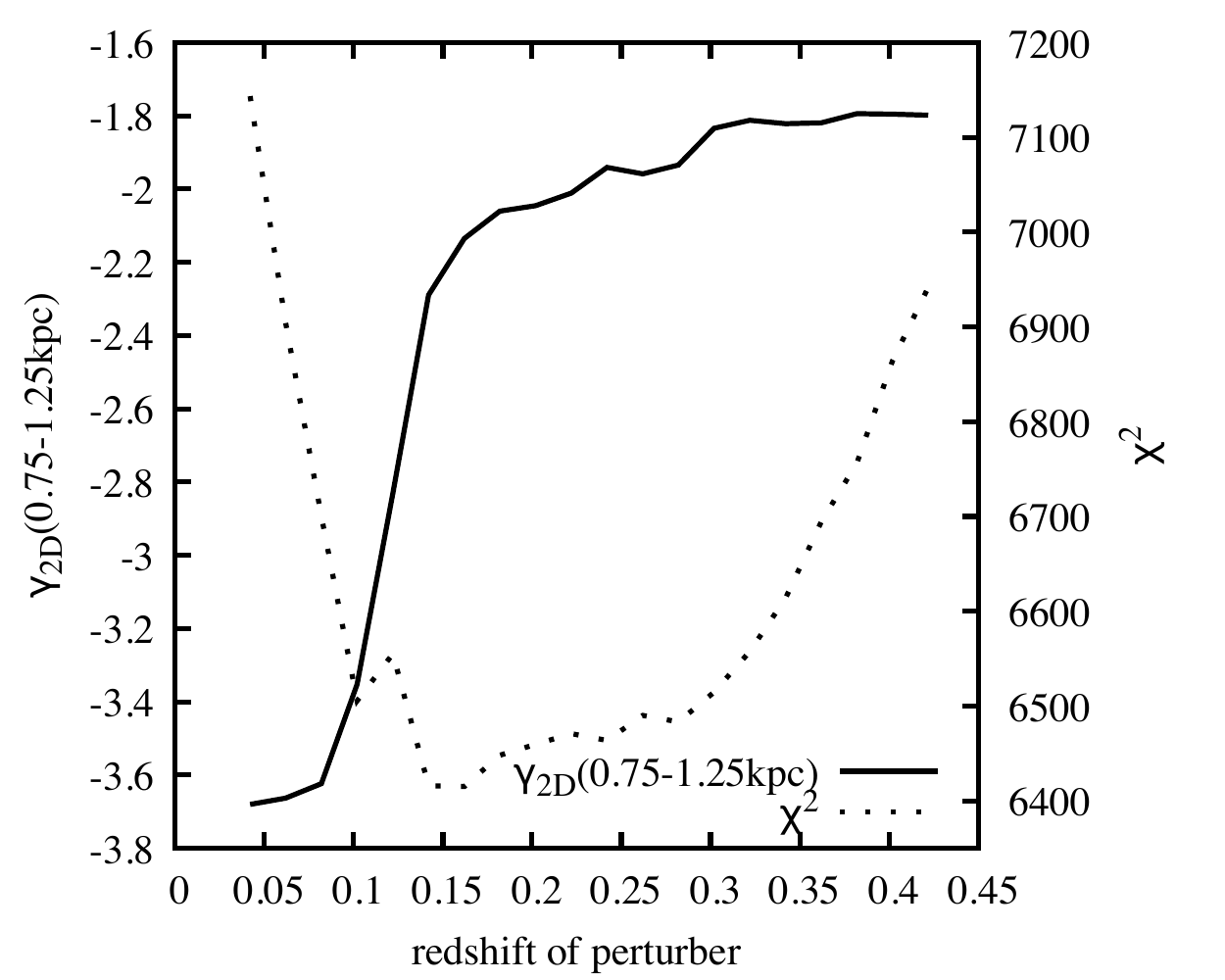}
		\label{fig:slope_z}
	}
	\caption{Best-fit perturber's projected mass within 1 kpc and average log-slope $\gamma_{2D}$(0.75-1.25kpc) as the perturber's redshift is varied. In (a) the $M_{2D}$(1kpc) is given by the solid curve, while the $\chi^2$ of the fit is given by the dashed curve with corresponding values given by the axis on the right. In (b) the log-slope is given by the (solid curve), again with the $\chi^2$ given by the dashed curve for comparison. These plots were generated by taking the best-fit model with truncated NFW profile, gradually varying the redshift in either direction with steps of $\Delta z = 0.02$ and reoptimizing the $\chi^2$ function.} \label{fig:m1kpc_slope_z}
\end{figure*}

Several studies have suggested that a significant fraction of lensing perturbations are caused by dark matter halos lying along the line of sight to the lensed source, instead of being subhalos of the lensing galaxy \citep{despali2018,li2017,xu2012}. Unfortunately in our fits, the tidal radius of the perturber is largely prior dominated and hence cannot be used to determine \emph{a priori} whether the perturber is indeed a subhalo or not. This raises the question, can the inferred density of the perturber in SDSSJ0946+1006 be reduced if it lies along the line of sight, such that the tension with CDM is mitigated? For example, if the perturber lies behind the lensing galaxy at z=0.4, its angular diameter distance is nearly 50\% higher compared to the lens redshift (z=0.222). Assuming the perturbation radius would be constrained to have a similar angular size ($\sim 0.3''$), then the same mass within this perturbation radius would now be spread over roughly 1.5 kpc in radius, perhaps implying a lower central density. On the other hand, if the perturber lies in front of the lens at lower redshift, its Einstein radius becomes larger, perhaps indicating the perturber does not need to be so massive. Either effect could possibly reduce the mass of the perturber contained within 1 kpc. In this section, we investigate whether the assumption of a line-of-sight perturber can lessen the tension with CDM.

The complicating factor in these scenarios is the role played by recursive lensing: instead of one lens plane, there are now two, and the lensing proceeds in two stages as the lensed light from the more distant galaxy is in turned lensed by the closer one \citep{schneider1992}. This recursive lensing has observable effects that can alter the quality of the fit, for better or worse. This is most obviously true when the perturber is behind the primary lens at high redshift, so that the convergence from the perturber is essentially lensed by the primary lens galaxy. This has the effect of smearing its convergence into an arc, or in extreme cases even leading to multiple ``images'' of the same perturber (i.e., perturbations in two different locations in the lens plane).

We now investigate whether changing the redshift of the perturber can lead to a less dense halo without significantly degrading the fit, taking into account the full effect of recursive lensing by solving the recursive lensing equation \citep{schneider1992}. In principle one could redo the analyses of Section \ref{sec:results}, but varying the perturber's redshift as an additional model parameter. However, the degeneracies between the subhalo parameters ($m_{200},r_s,r_t,z_{pert}$) would require running the nested sampling algorithm for a much longer time to achieve convergence, straining our computational resources. We thus opt for a simpler investigative approach: starting from the best-fit model for the subhalo, we gradually change the redshift of the perturber, by steps $\Delta z = 0.02$, reoptimizing the chi-square at every step using Powell's method to find a new set of best-fit lens and source parameters. First we reduce the redshift down to $z=0.05$, bringing the perturber in front of the lens plane; the procedure is then repeated going in the other direction up to $z=0.45$. Provided that the redshift is changed slowly enough, one is still tracing out the approximate best-fit model as a function of redshift.

The results are plotted in Figure \ref{fig:m1kpc_slope_z} for the tNFW model (the corresponding plots for the tNFWmult model whose behavior is qualitatively similar, are given in Appendix \ref{sec:appendix_los}, Figure \ref{fig:m1kpc_slope_z_mult}). In Figure \ref{fig:m1kpc_slope_z}(a) we plot the perturber's projected mass within 1 kpc (solid line, left axis) along with the best-fit $\chi^2$ value (dashed line, right axis) as a function of the perturber's redshift. Surprisingly, Figure \ref{fig:m1kpc_slope_z} shows the the variation in goodness-of-fit is equally apparent at both high and low redshifts.  The recursive lensing effect is entirely responsible for this variation in the $\chi^2$: if the recursive lensing effect is ``turned off'', so that the convergences are simply added together, one can always achieve an equally good chi-square by rescaling the masses and scale lengths regardless of redshift. Hence, the fact that the $\chi^2$ increases at low redshifts shows that the recursive lensing effect is not only important behind the lens plane, but also in front as the redshift gets lower. This is because the perturber is strongly lensing by itself (with Einstein radius of $\sim 0.1''$) and its Einstein radius increases as it comes closer, for fixed $m_{200}$ and $r_s$ values.\footnote{For completeness' sake, we note that if you bring the perturber much closer than z=0.05, its Einstein radius will finally decrease again as the mass "spreads out" over the lens plane, driving the surface density per arcsec to be low. But the perturber in this case is so dense that this effect has not kicked in yet by z=0.05.} Hence, to achieve a similar perturbation radius (in angular size), the mass of the perturber must be reduced, driving the mass within 1 kpc to be lower for z=0.05, but by this point, the fit is degraded significantly, with the $\chi^2$ increasing by $\sim 500$. (See Appendix \ref{sec:appendix_los} for a discussion of how the perturbation size varies with redshift.)

Although one achieves a similar (or perhaps slightly better) fit down to $z_{pert}=0.15$, the mass scale $M_{2D}$(1kpc) is only slightly reduced. To get down to 1 $M_\odot$ within 1 kpc, the perturber must be brought down to $z=0.05$, and the fit is quite poor by this time with significant residuals in the vicinity of the subhalo. In addition, Figure \ref{fig:m1kpc_slope_z}(b) shows that the log-slope $\gamma_{2D}$(1kpc) becomes extremely steep in this limit, diving below -3.6 at $z=0.05$, due to the best-fit tidal radius becoming very small. This indicates an extremely high concentration (indeed, approaching the limit of a point mass) and the slope is much steeper than that of field halos in CDM, as Figure \ref{fig:TNG_subs_contours_cosmoRels} shows (black curves give the log-slope for concentrations $3\sigma$ above the median value at different redshifts). Therefore the limit of low redshift, while leading to a lower $M_{2D}$(1kpc), also exacerbates the concentration issue and leads to a significantly poorer fit.\footnote{A related question is whether the perturber could be a black hole within the Milky Way. We did a PolyChord run with a point mass perturber (and no multipoles) at the redshift of the lens and found a best-fit solution with $\chi^2$/pixel $\approx 1.71$, only modestly worse than the tNFW or CoreCusp models listed in Table \ref{tab:models}. Following the above procedure, if we gradually reduce the perturber redshift down to $z=10^{-6}$ (about 4 kpc away), the pertuber has a best-fit mass $\approx 1.1\times10^4M_\odot$, consistent with a large intermediate-mass black hole. However, there are strong residuals in the vicinity of the perturber and we have $\chi^2$/pixel $\approx 1.95$, indicating the fit is severely degraded in the low-redshift limit, consistent with the results using tNFW in this section.}

We conclude that the assumption of a line-of-sight perturber does not significantly lessen the tension with CDM; indeed, it likely worsens the tension, since field halos are in general less dense than subhalos of similar mass. In addition, even if a slightly better fit might be achieved down to $z_{pert}=0.15$, the upper bound on the stellar luminosity of the perturber becomes stricter: the luminosity distance at $z=0.15$ is less than 70\% its value at the primary lens redshift, which leads to the upper bound on luminosity being less than half as bright, making it harder to explain with a very massive halo. Thus, although both scenarios are in tension with CDM, we consider the scenario of a perturbing subhalo to be more plausible than that of a line-of-sight halo.

\section{Discussion: Can dark matter physics explain the high concentration of the subhalo in lens SDSSJ0946+1006?}\label{sec:sidm}

Having established that the subhalo in SDSSJ0946+1006 is an outlier in CDM based on the lensing constraints, we ask if a modification of the dark matter particle physics can explain its high central density and concentration? Alternatives to cold, collisionless CDM (such as warm dark matter or SIDM) are typically invoked to mitigate the cusp-core or ``too big to fail'' problems by \emph{reducing} the central densities of small dark matter halos \citep{Vogelsberger:2012MNRAS.423.3740V,lovell2014}, which would appear to exacerbate the problem here. 

However, one intriguing possibility is the phenomenon of core collapse due to dark matter self-interactions \citep{ahn2005,elbert2014}. In SIDM, the self-interactions between dark matter particles thermalize the central regions of dark matter halos, producing a constant density core \citep{Dave:2000ar}. Over long enough timescales, however, the central core collapses as heat is transferred outward (the ``gravothermal catastrophe'' familiar in globular clusters; \citealt{quinlan1996}), resulting in high central densities. 
For subhalos, the collapse timescale can be significantly shorter than for field halos \citep{nishikawa2020}. It has been argued that this phenomena could be playing a role in setting the observed diversity of dark matter densities measured in the Milky Way satellite galaxies \citep{kahlhoefer2019,Zavala:2019sjk,kaplinghat2020,Sameie:2019zfo,Turner:2020vlf}. Could this core collapse phenomenon also be responsible for the observed properties of the substructure in SDSSJ0946+1006? It is not possible to answer this question at present. The interplay between the velocity dependence of the cross section, the orbit of the subhalo, the concentration at infall and the tidal field of the host halo allow for a rich variety of evolution in the subhalo density \citep{kahlhoefer2019}, which needs to be investigated in detail. We highlight in passing another intriguing possibility that core collapse could also explain the apparent high concentration of the dark matter substructure discovered in the GD-1 stellar stream by \cite{bonaca2019}.

\section{Conclusions}\label{sec:conclusions}

We have fit several lens models to the gravitationally lensed images in SDSSJ0946+1006 and confirm the discovery of a dark and remarkably concentrated subhalo made by \cite{vegetti2010}. Our most robust measurement is the subhalo's projected mass within 1 kpc of its center of $(2 - 3.7)\times 10^9M_\odot$, with our highest evidence model inferring $M_{2D}$(1kpc) = $(3.3\pm0.3)\times 10^9 M_\odot$. This mass inference is robust to changes in the subhalo's assumed density profile or tidal radius, as 1 kpc is quite close to the subhalo's lensing perturbation radius defined in \cite{minor2017}. This is an extraordinarily high central density given our conservative upper bound on its stellar luminosity $L_V \approx1.2\times10^8L_\odot$. In addition to the mass, we find the average log-slope of the subhalo's projected density profile over the range 0.75-1.25 kpc to be steeper than isothermal for all models, with the highest-evidence model giving $\gamma_{2D} = -1.27_{-0.13}^{+0.11}$. Our reconstructions of the lensed images and source galaxy when the subhalo is modeled with a truncated NFW profile are shown in Figure \ref{fig:bestfit} and \ref{fig:source}, and the inferred lens parameters for all models are given in Table \ref{tab:models}.

To test whether such a dense subhalo is expected to occur in $\Lambda$CDM, we compared our inferred subhalo mass and density slope to those of subhalos within 167 analogue lensing galaxies in the Illustris TNG100-1 hydrodynamical simulation, and 188 analogue lensing galaxies in the Illustris TNG100-1-DARK DMO simulation. The results are encapsulated in Figures \ref{fig:TNG_subs_contours_cosmoRels} and \ref{fig:cumdist_slope_34}. We conclude with the points listed below. 

$\bullet$ By generating mock ``observations'' along many lines of sight for each subhalo in our Illustris TNG100 sample, we find that while many subhalos have projected masses within 1 kpc that satisfy the lensing constraint, fewer than 1\% of such observations produce projected density slopes consistent with the lensing constraints at the 95\% confidence level (Figure \ref{fig:cumdist_slope_34}). This implies that if a CDM subhalo produces a lensing perturbation of similar angular size to what is observed in SDSSJ0946+1006, the likelihood of the subhalo having a density slope as steep as our lensing constraints is $<1\%$. This conclusion can be made independently of the subhalo's stellar mass.

$\bullet$ Among the simulated subhalos that have a projected mass within 1 kpc of $>2\times10^9M_\odot$ along at least one line of sight, as required by the lensing constraints, \emph{all} of them have stellar masses exceeding our conservative upper bound for the observed subhalo ($M_* < 2.4\times10^8M_\odot$) for which we assume a stellar mass-to-light ratio $M_*/L_V = 2M_\odot/L_\odot$, consistent with a Salpeter-like IMF. This is the case regardless of the subhalo's density slope (i.e. concentration). For the actual subhalo to have a stellar mass consistent with the TNG100 candidates would require at least $M_*/L_V \gtrsim 8M_\odot/L_\odot$, which would be unprecedentedly high in a dwarf galaxy.

$\bullet$ 
The small subset of simulated subhalos that are consistent with the lensing constraints (both projected mass and density slope) within 99\% CL for specific lines of sight all have very high stellar masses within 1 kpc ($\gtrsim 3\times10^9M_\odot$, higher than the dark matter mass). 
These subhalos are likely to have had their central dark matter fractions boosted by adiabatic contraction, a view that is supported by the fact that \emph{none} of the mock observations in the DMO simulation achieve a steep enough slope (red curve in Figure \ref{fig:cumdist_slope_34}). In view of our upper bound on the subhalo's stellar mass ($2.4\times10^8M_\odot$), adiabatic contraction is unlikely to have been a significant factor in the observed subhalo.


$\bullet$ We find that the tension with CDM cannot be mitigated if the perturber is a field halo along the line of sight, rather than a subhalo of the lens galaxy: although the perturber's best-fit projected mass within 1 kpc drops below $2\times10^9M_\odot$ for a low-redshift perturber, the goodness-of-fit plummets along with it, and the best-fit density slope becomes much steeper (Figure \ref{fig:m1kpc_slope_z}). In addition, the constraint on the perturber's luminosity becomes more stringent at low redshift, requiring an even smaller stellar mass, while the lack of tidal stripping in the field makes the density slope harder to explain. 

Thus, the central dark matter density, the concentration and the apparent darkness of the observed subhalo in  SDSSJ0946+1006 make it an extreme outlier in the $\Lambda$CDM model. 


Perhaps the most plausible way to reconcile the subhalo with CDM is if a lensing solution could be found with projected mass $M_{2D}$(1kpc) $\lesssim 2\times10^9M_\odot$ within 1 kpc, since subhalos in this range can have a stellar mass consistent with our upper bound ($M_* < 2.4\times10^8M_\odot$). Since $M_{2D}$(1kpc) is robust to changes in the subhalo model, such a solution would likely require a more sophisticated model for the host galaxy. However, our lensing solutions that achieve smaller $M_{2D}$(1kpc) require significantly steeper slopes (Figure \ref{fig:TNG_subs_contours_cosmo_corecusp}), worsening the tension with CDM, so it is unclear whether such a solution would be viable. Fitting both of the lensed sources in SDSSJ0946+1006 with a more flexible host galaxy model (along with the subhalo) may provide clarity on this question, perhaps in combination with spatially resolved kinematic data for the lens galaxy. In addition, repeating our analysis in Illustris TNG50 would also provide a more robust measurement of the subhalos' dark matter slopes along individual lines of sight, particularly at lower subhalo masses.



If the unexpected properties of the perturbing subhalo in SDSSJ0946+1006 
are due to the particle physics of dark matter, such as dark matter self-interactions, one may expect that many more perturbers of its kind will be detected among the thousands of strong lenses expected to be discovered by the Euclid and LSST surveys. Such highly concentrated subhalos may also be detected by their perturbations of tidal streams; indeed, such a detection may have already occurred \citep{bonaca2019}. If so, we may be catching our first glimpses of a population of low-mass, concentrated dark matter subhalos whose properties would allow us to directly constrain the particle physics of dark matter in the coming years.

\section*{Acknowledgements}

We would like to thank Giulia Despali, Anna Nierenberg, David Zurek, and Mordecai Mac-Low for insightful discussions during the course of the project. We thank Alessandro Sonnenfeld for supplying multiband data to help test our lensing reconstructions. QM was supported by NSF grant AST-1615306 and MK by NSF PHY-1915005.
SV has received funding from the European Research Council (ERC) under the European Union’s Horizon 2020 research and innovation programme (grant agreement No 758853).
We gratefully acknowledge a grant of computer time from XSEDE allocation TG-AST130007.

This research was also supported, in part, by a grant of computer time from the City University of New York High Performance Computing Center under NSF Grants CNS-0855217, CNS-0958379 and ACI-1126113.

\bibliography{subhalo3}

\begin{appendix}
\setcounter{section}{0}
\section{A. Candidate Subhalos from Illustris TNG100-1}\label{sec:appendix_candidates}

\begin{figure}
	\centering
	\includegraphics[height=0.75\hsize,width=0.99\hsize]{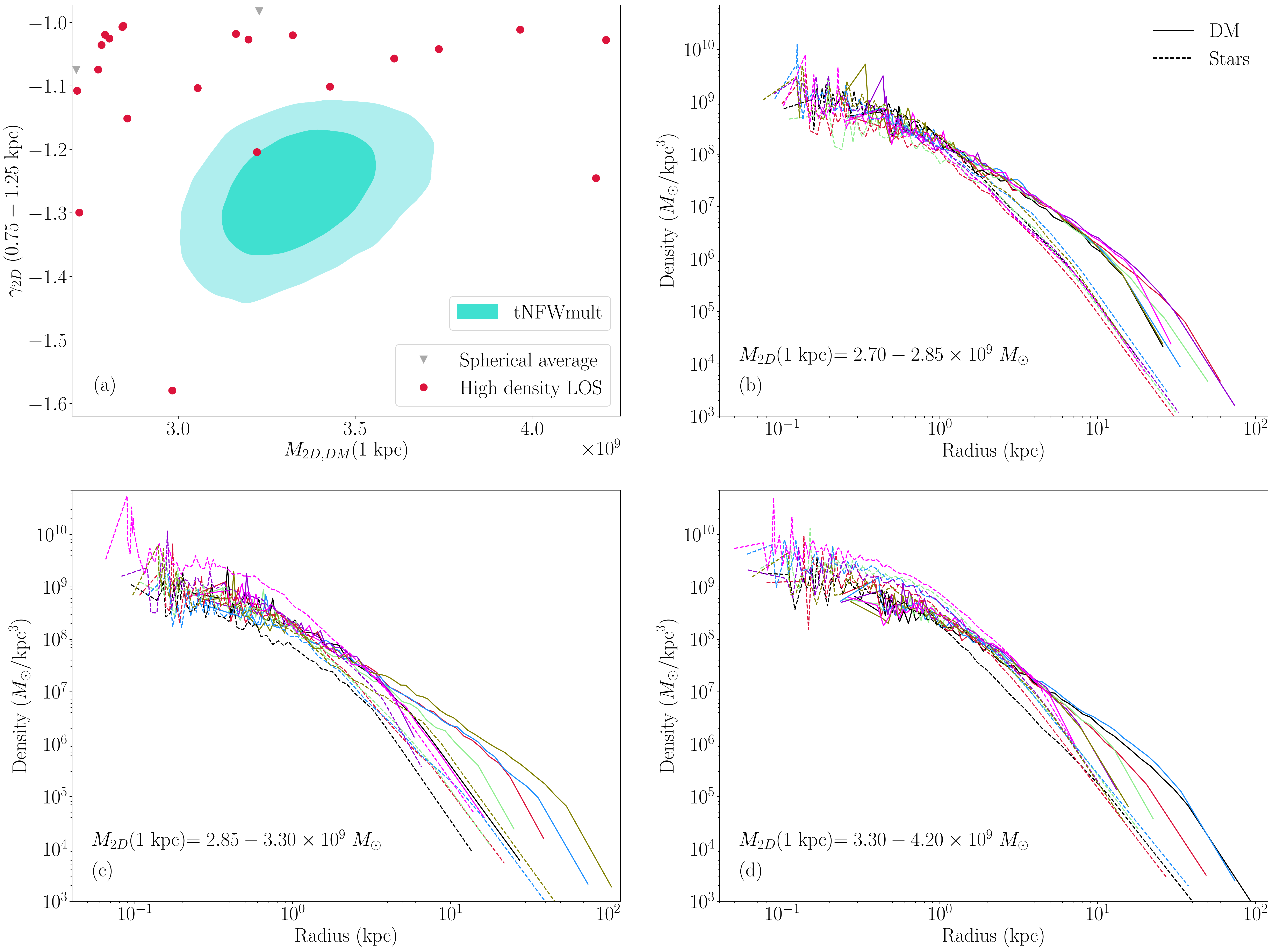}
	\caption{The 21 TNG100-1 subhalo candidates in closest range to the lensing constraints from the tNFW model with multinest (panel (a)). 3D dark matter (solid line) and stellar (dashed line) density profiles corresponding to the subhalo candidates in panel (a), arranged by increasing projected mass in 1 kpc (panels (b),(c),(d)).}
\label{fig:TNG_candidates_density}
\end{figure}

To take a closer look at the closest subhalo candidates in TNG100-1, we further investigated 21 subhalos that had the closest match with the lensing constraints by selecting subhalos with $2.7\times10^9 M_\odot < M_{2D,DM}(1\mathrm{kpc}) < 4.2\times10^9 M_\odot$ and $\gamma_{2D} < -1$. These subhalos are shown in panel 1 of Figure \ref{fig:TNG_candidates_density}. Notice here that only two of these subhalos make the cut when full spherical averaging was used. For these subhalos, we plot and show the 3D stellar and dark matter density profiles, grouping them by increasing projected mass within 1 kpc (panels 2, 3, and 4 in Figure \ref{fig:TNG_candidates_density}). In all the panels, the stellar densities are either higher than the dark matter densities or are comparable near the central region of the subhalo. The subhalos with the highest projected mass in 1 kpc (panel 4 in Figure \ref{fig:TNG_candidates_density}) have stellar densities that are higher than their respective dark matter densities by as much as an order of magnitude. These high stellar densities strongly indicate that adiabatic contraction may have modified the dark matter halo structures of these subhalos, steepening them further than the NFW expectation. We thus conclude that the subhalo analogs found in TNG100-1 are analogs because AC steepens their slopes and/or pushes enough mass into 1 kpc, combined with tidal truncation during the subhalo's pass through pericenter along its orbit in its host halo.


\section{B. Comparison of all four lensing models to Illustris TNG100-1 sample}\label{sec:appendix_lensmodels}

Here we make a comparison of all four of our lensing models (whose inferred parameters are given in Table \ref{tab:models}) to the subhalo sample in the IllustrisTNG simulation presented in Section \ref{sec:illustris}.

\begin{figure}
	\centering
	\includegraphics[height=0.4\hsize,width=1.02\hsize]{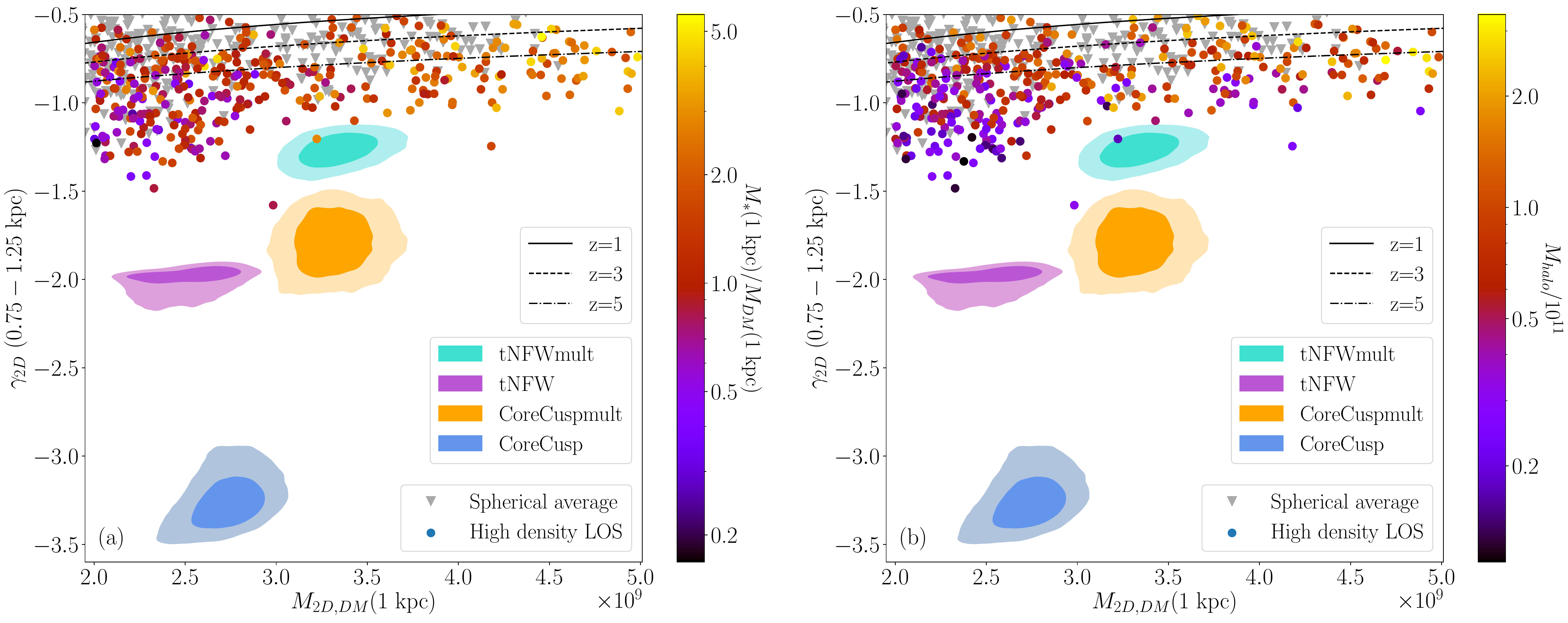}
	\caption{The same as Figure \ref{fig:TNG_subs_contours_cosmoRels}, except here we plot the contours for all four models: tNFW and core-cusp, both the elliptical case as well as with multipoles. Here the points from the high density LOS average are color coded by $M_*(1 \mathrm{kpc})/M_{DM}(1 \mathrm{kpc})$ in panel (a), and by total halo mass in panel (b).}
\label{fig:TNG_subs_contours_cosmo_corecusp}
\end{figure}

\begin{figure*}
	\centering
	\subfigure[$2\times10^9 < M_{2D}$(1kpc)$/M_\odot < 3\times10^9$]
	{
		\includegraphics[height=0.38\hsize,width=0.47\hsize]{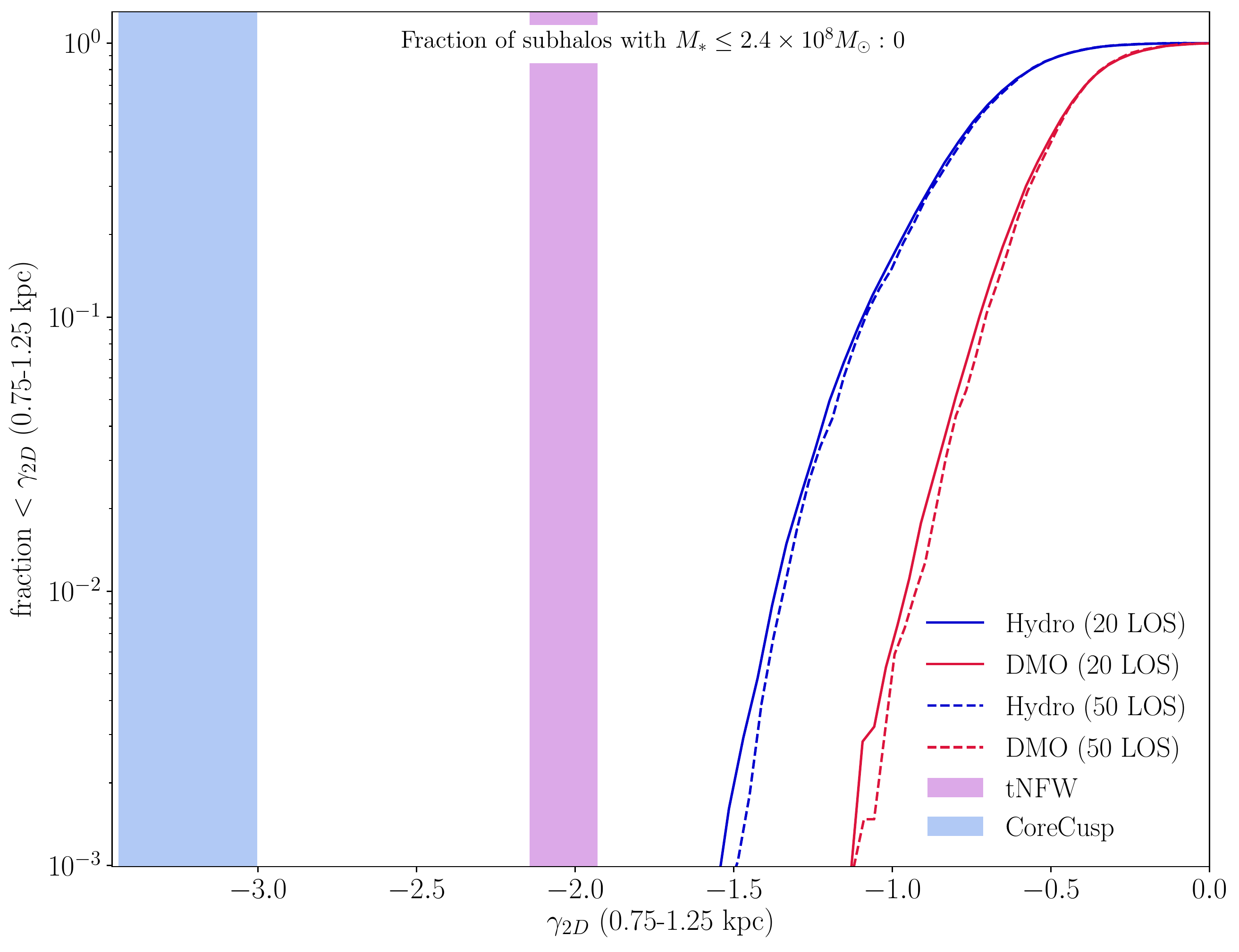}
		\label{fig:cumdist_23_allmodels}
	}
	\subfigure[$3\times10^9 < M_{2D}$(1kpc)$/M_\odot < 4\times10^9$]
	{
		\includegraphics[height=0.38\hsize,width=0.47\hsize]{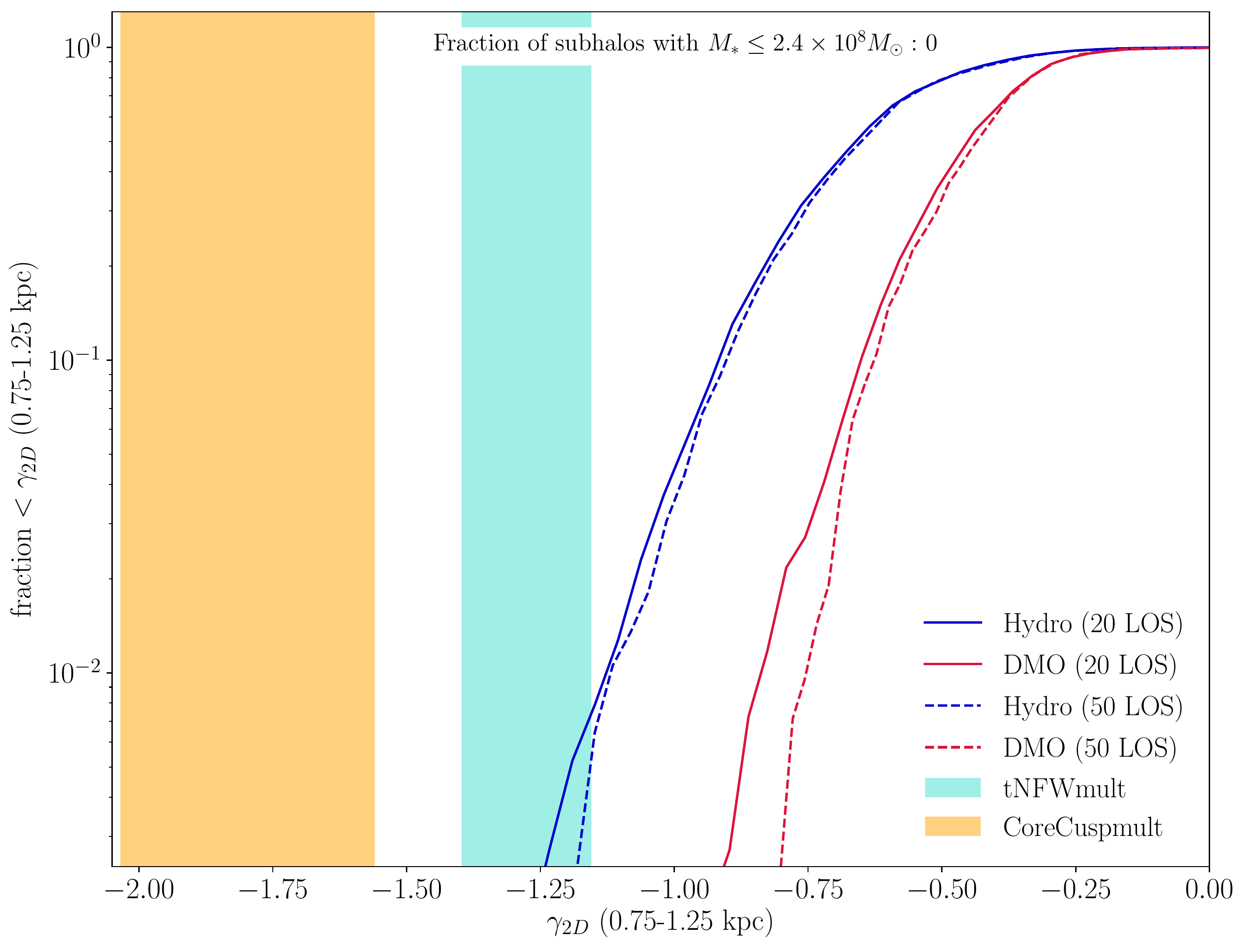}
		\label{fig:cumdist_34_allmodels}
	}
	\caption{The same as Figure \ref{fig:cumdist_slope_34}, except we plot two different ranges in $M_{2D}$(1kpc): figure (a) plots the 2-3$\times10^9M_\odot$ range favored by the elliptical tNFW and corecusp models, while figure (b) plots the 3-4$\times10^9M_\odot$ range favored by the models that include multipoles (tNFWmult and CoreCuspmult). Vertical bars respresenting the 95\% credible interval for each lens model are overlaid. We plot the cumulative distribution in $\gamma_{2D}$ for two different levels of averaging: for the solid curves, each mock ``observation'' is produced by averaging the profiles for 20 nearby lines of sight (out of 1000 total for each subhalo), while for the dashed curves, 50 LOS are averaged. Note that less than 0.1\% of observations that produce density slopes consistent with the lensing constraints for all lens models with the exception of tNFWmult, our most conservative model, for which the fraction is $<1\%$.}
\label{fig:cumdist_allmodels}
\end{figure*}

In Figure \ref{fig:TNG_subs_contours_cosmo_corecusp} we make a plot that is identical to Figure \ref{fig:TNG_subs_contours_cosmoRels}, except that all four of our lens models are shown. Note that the models with elliptical host galaxy (tNFW and corecusp) both have consistent subhalo masses within 1 kpc, while the models that include multipoles (tNFWmult and CoreCuspmult) have nearly identical inferred masses within 1 kpc. This is a consequence of the fact that the subhalo's perturbation radius (defined by the distance of the subhalo's center to the point of maximum warping of the lens' critical curve), is nearly 1 kpc, and hence the projected mass within 1 kpc is independent of the subhalo's density profile \citep{minor2017}. However, the tNFW, CoreCuspmult and CoreCusp models all infer significantly steeper slopes $\gamma_{2D}$(0.75-1.25kpc) compared to the tNFWmult model, and hence are even more difficult to explain in IllustrisTNG-100. 

Next, we examine the likelihood of the simulated subhalos having a projected mass and density slope consistent with the model. Using the same procedure for generating mock ``observations'' as in Section \ref{sec:illustris_slope_pct}, we plot the cumulative distributions in $\gamma_{2D}$. The results are shown in Figure \ref{fig:cumdist_allmodels}. We now examine two different ranges in $M_{2D}$(1kpc): 2-3$\times10^9M_\odot$, favored by the tNFW and CoreCusp models, and 3-4$\times10^9M_\odot$, favored by the models with multipoles (tNFWmult, CoreCuspmult). Vertical bars show the 95\% credible interval (defined by the 2.5\% and 97.5\% percentiles of the marginalized posterior in $\gamma_{2D}$) for each model. To examine whether our results are dependent on how we average the profiles over nearby lines of sight, we plot the distributions for 20 LOS averaging (solid curve) and 50 LOS averaging (dashed curve) for both the hydro (blue) and DMO (dashed) simulations. Note the resulting curves are relatively unchanged whether we average over 20 or 50 LOS, although the 50 LOS averaging gives a less smooth distribution due to having fewer observations overall. Remarkably, in the tNFW, CoreCusp and CoreCuspmult models, none of all observations give a steep enough slope to be consistent with the lensing constraints. For our most conservative model, tNFWmult (which is the one shown in Figure \ref{fig:cumdist_slope_34}), the fraction is $<1\%$. The fractions become negligible in the DMO sample for all models. We conclude from this that the actual subhalo is an outlier in its density slope (and hence concentration) in CDM for all models. In addition, as discussed in Section \ref{sec:discussion}, the few subhalos that do achieve a steep enough slope for some lines of sight all have $>10^9M_\odot$ within 1 kpc, far exceeding our conservative bound on the subhalo's stellar mass.

From the lens modeling point of view, one might hold out hope that a more sophisticated model could yield a solution that is more in accord with $\Lambda$CDM predictions--either by reducing $M_{2D}$(1kpc), or by yielding a shallower slope. However, as discussed in Section \ref{sec:rpert}, there may be relatively little wiggle room for this (a third possibility, that the perturber is a line-of-sight halo rather than a subhalo, was discussed in Section \ref{sec:los}). The most robust inference that can be made from the subhalo is its perturbation radius and the subhalo's projected mass enclosed therein. The four models we fit to the subhalo yielded perturbation radii whose values (roughly 1 kpc) differ by as much as a pixel length, as one might expect; it is this difference that results in different inferred masses within 1 kpc, spanning a range 2-3.5$\times 10^9M_\odot$.

\begin{figure*}
 	\centering
 	\includegraphics[height=0.6\hsize]{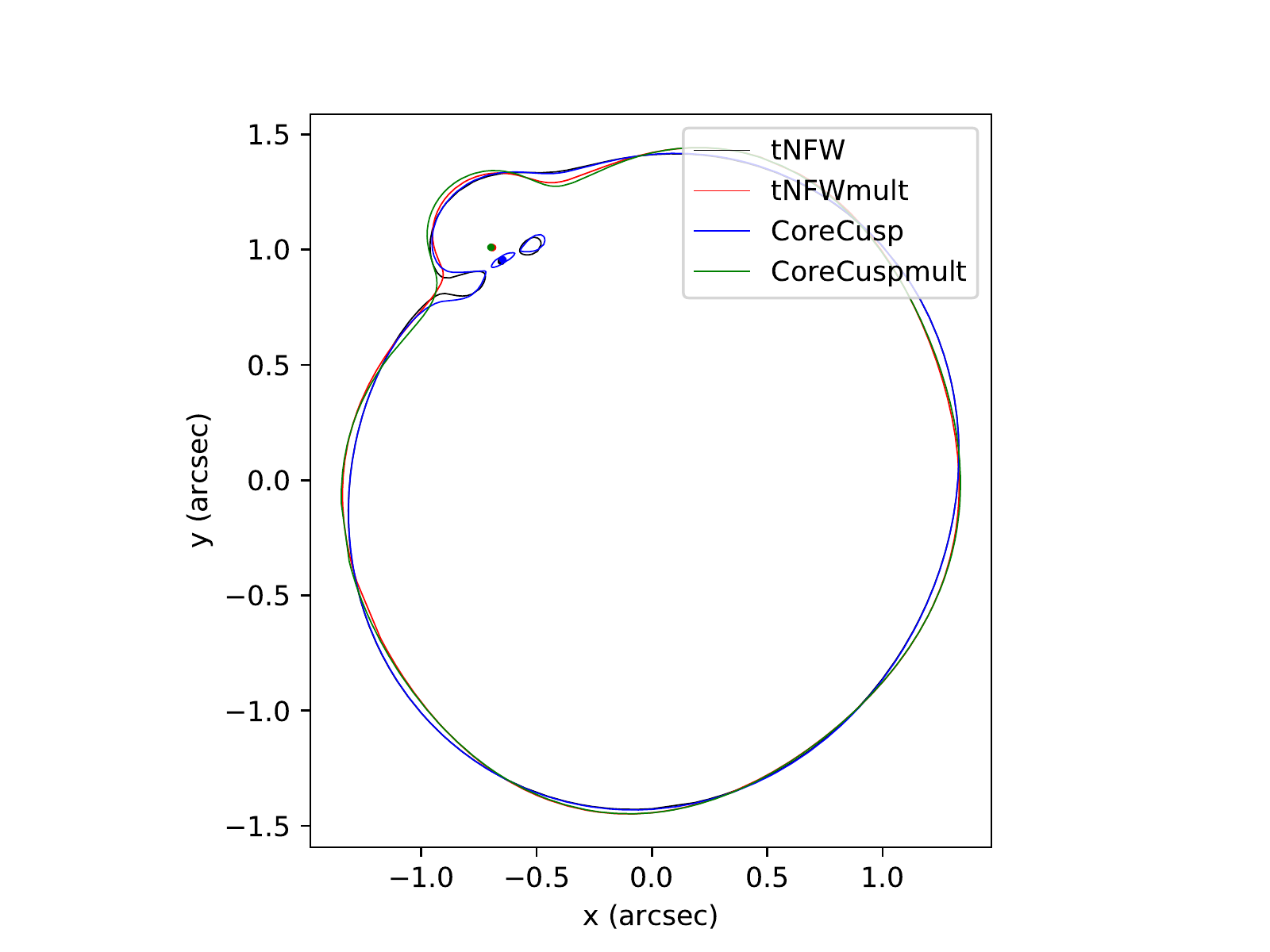}
 	\caption{Critical curves for all four lens models we fit to SDSSJ0946+1006, whose inferred parameters are listed in Table \ref{tab:models} The best-fit position of the subhalo's center for each model is shown by the filled circles, whose colors match the corresponding critical curve colors. Note the size of the subhalo's perturbation to the critical curve (in the upper left) is fairly consistent in all four models. However, the position of the subhalo differs noticeably in the models that include multipoles compared to those that do not; as a result, the inferred perturbation radius differs by up to roughly 0.05'', the width of an image pixel.}
 \label{fig:critical_curves}
 \end{figure*}

\begin{figure*}
	\centering
	\includegraphics[height=1.0\hsize]{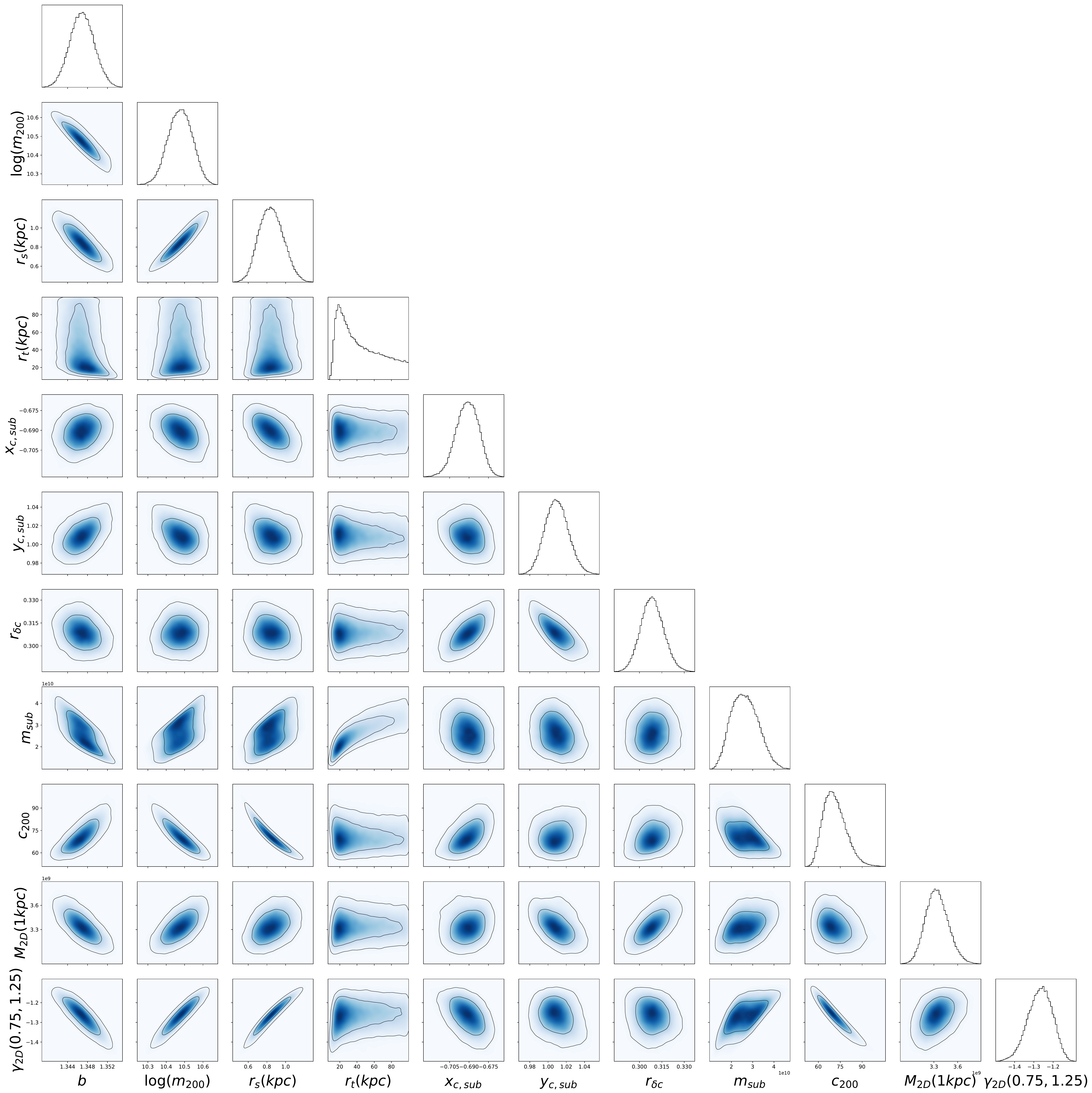}
	\caption{Posteriors in the subhalo lensing parameters for model tNFWmult, whose uncertainties are tabulated in Table \ref{tab:models}. The host galaxy parameters are omitted here except for $b$, the Einstein radius parameter. The structural parameters of the subhalo $r_s$, $r_t$ are given in kpc, while the position coordinates and perturbation radius $r_{\delta c}$ are given in arcseconds.}
\label{fig:posts}
\end{figure*}

In Figure \ref{fig:critical_curves} we plot the critical curves from all four models, along with the subhalo position in each model; note that all four models have a critical curve perturbation of similar size, but differ slightly in the subhalo position (resulting in a slightly different $r_{\delta c}$). The covariance between subhalo position and perturbation radius is evident in the posteriors. For example, in Figure \ref{fig:posts} we plot joint posteriors in the subhalo parameters for the tNFWmult model. The joint posteriors in $r_{\delta c}$ versus $x_{c,sub}$ and $y_{c,sub}$ show that if the subhalo is moved toward higher $x$ and/or lower $y$, $r_{\delta c}$ increases; the same relationship is evident in all the models we fit. A similar covariance with the center coordinates is seen with $M_{2D}$(1kpc) as well. However, as all models agree on the size of the critical curve perturbation itself (Figure \ref{fig:critical_curves}), we consider it unlikely that a model could produce a much smaller $r_{\delta c}$ without degrading the fit. This is supported by the results of Section \ref{sec:los}: when the perturber is modeled as a field halo along the line-of-sight instead of a subhalo, for sufficiently low redshift $(z \lesssim 0.1)$ the goodness-of-fit plummets as the best-fit $M_{2D}$(1kpc) drops below $2\times10^9M_\odot$ (see Figure \ref{fig:m1kpc_slope_z}).

\section{C. Variation of best-fit subhalo mass and perturbation radius with redshift for tNFW, tNFWmult models}\label{sec:appendix_los}

\begin{figure*}
	\centering

	\subfigure[perturber mass within 1 kpc]
	{
		\includegraphics[height=0.38\hsize,width=0.48\hsize]{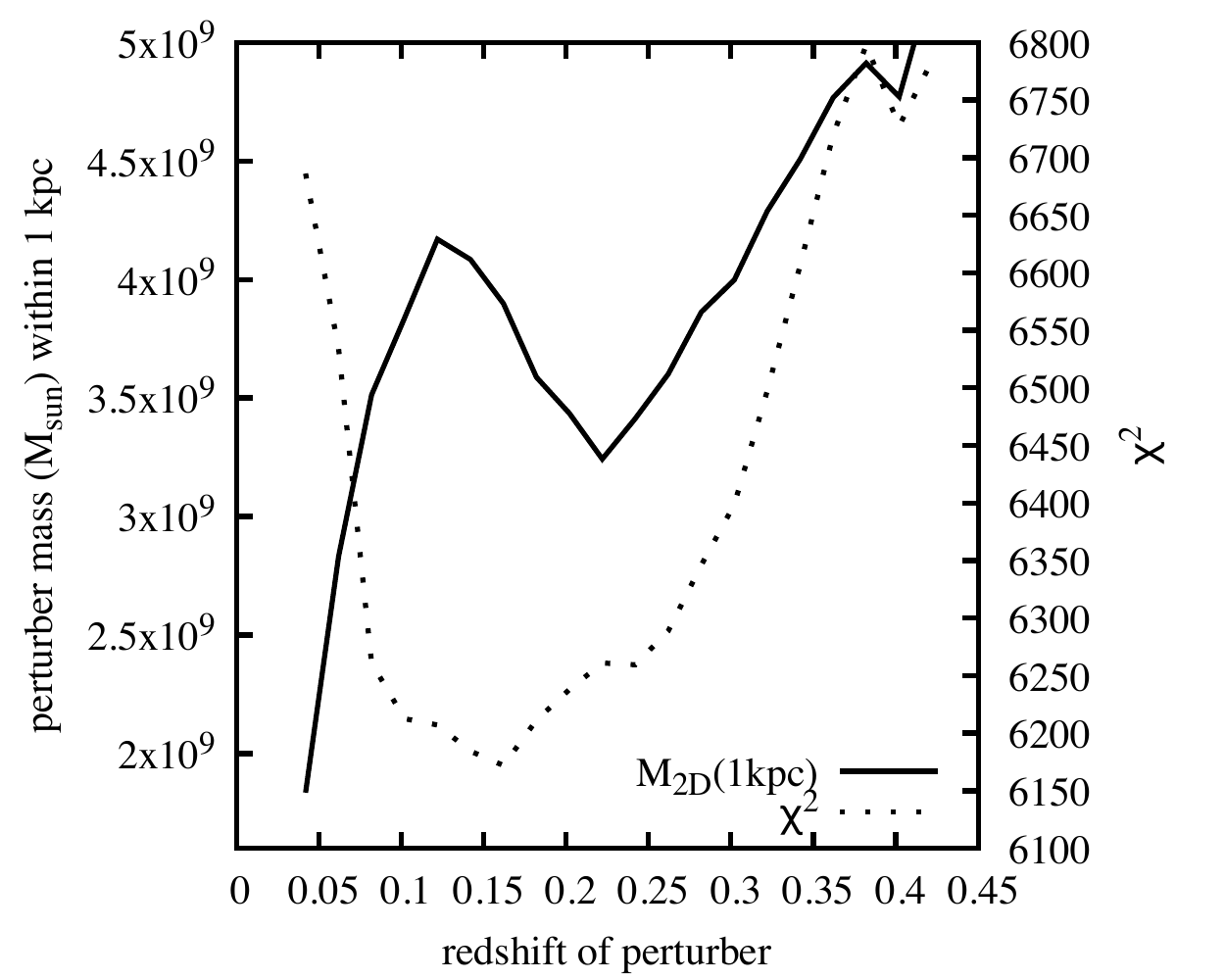}
		\label{fig:m1kpc_z_mult}
	}
	\subfigure[perturber log-slope]
	{
		\includegraphics[height=0.38\hsize,width=0.48\hsize]{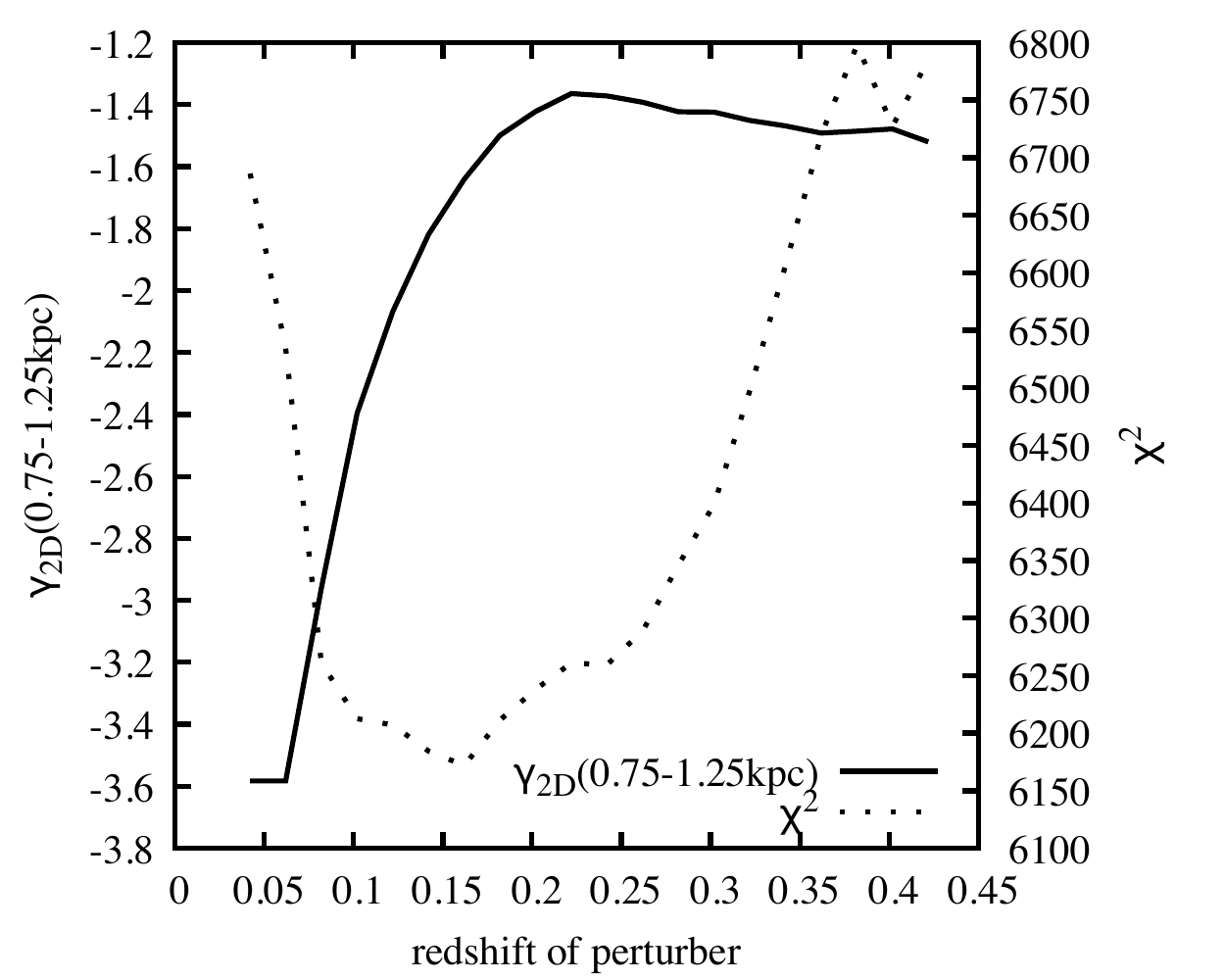}
		\label{fig:slope_z_mult}
	}
	\caption{Same as Figure \ref{fig:m1kpc_slope_z}, except using the tNFWmult model instead of tNFW. Note that as with the tNFW model, the log-slope becomes very steep in the limit of low redshift. Likewise, the perturber mass drops below $2\times10^9M_\odot$ for low redshift, but the fit becomes significantly degraded in this limit.} \label{fig:m1kpc_slope_z_mult}
\end{figure*}

\begin{figure}
	\centering
	\includegraphics[height=0.45\hsize,width=0.59\hsize]{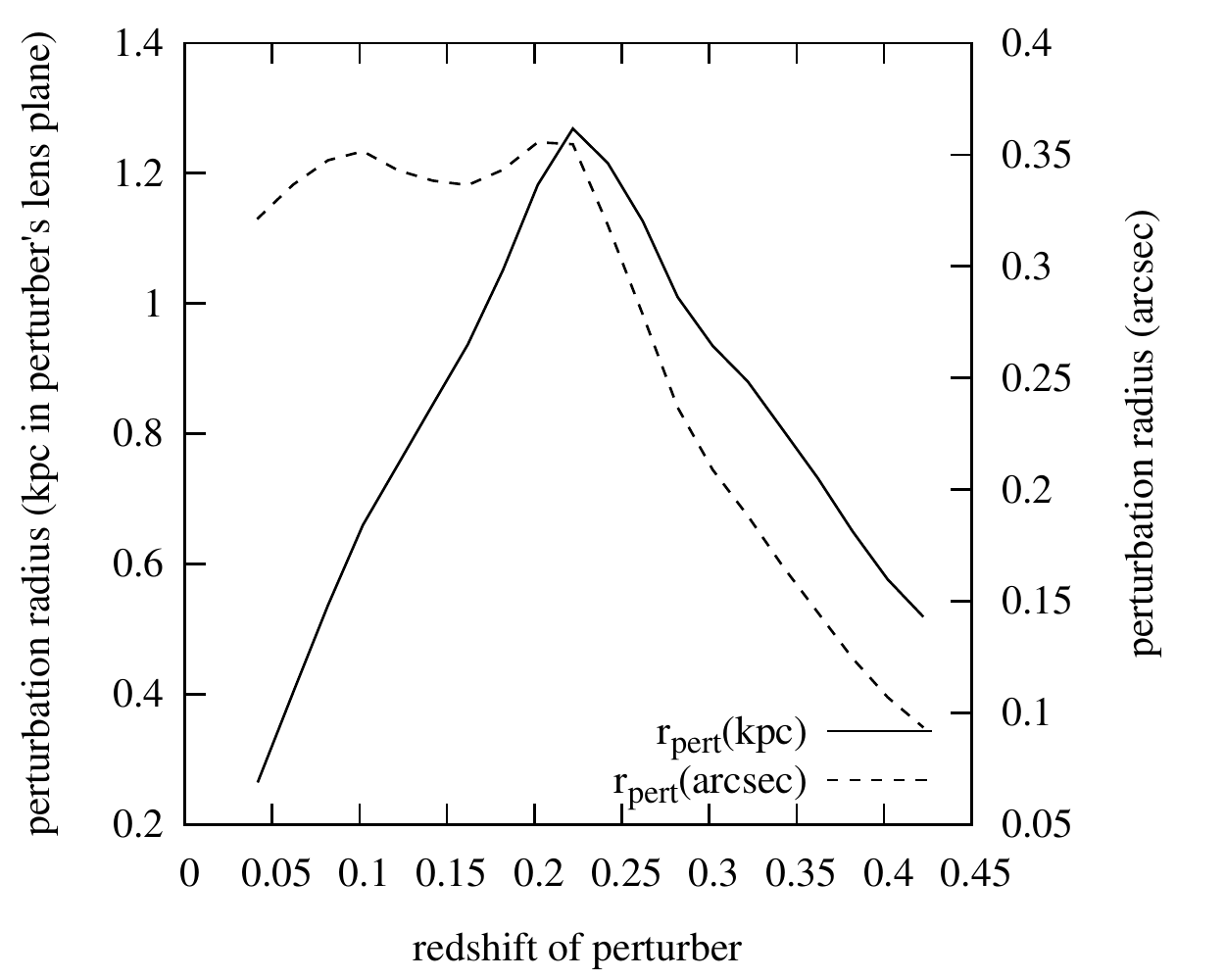}
	\caption{The perturber's perturbation radius (defined in Section \ref{sec:rpert}) as a unction of redshift. The perturbation radius is given in kpc (solid curve) and arcsec (dashed curve, axis on right). These plots were generated by taking the best-fit model with truncated NFW profile, gradually varying the redshift in either direction with steps of $\Delta z = 0.02$ and reoptimizing the $\chi^2$ function.}
\label{fig:rpert_z}
\end{figure}

In Figure \ref{fig:m1kpc_slope_z_mult} we plot the $M_{2D}$(1kpc) and $\chi^2$ as the perturber redshift is varied, starting from the tNFWmult model. Note that, as in Figure \ref{fig:m1kpc_slope_z}, the fit becomes significantly degraded as the perturber mass drops down to $2\times10^9M_\odot$ for low redshift. Likewise, the required slope becomes very steep at low redshift.

It is interesting to consider how the perturbation radius of the subhalo in SDSSJ0946+1006 changes as a function of redshift, since this has some bearing on whether $M_{2D}$(1kpc) can be considered a robust quantity.  In Figure \ref{fig:rpert_z} we plot the perturbation radius in units of kpc (solid line, left axis) and arcsec (dashed line, right axis). Note that at low redshift, the angular size of the perturbation remains approximately the same, while the physical size becomes smaller. This is easily understood, since the angular diameter distance becomes smaller as the redshift is reduced. Naively one might expect the mass within the (angular) perturbation radius is preserved, and hence there would be a greater overall mass within 1 kpc since this corresponds to a bigger angular size at low redshift. However, as discussed above, the lensing efficiency of the perturber increases at lower redshift, requiring a smaller mass to keep the same perturbation scale, and this is the dominant effect, as Figure \ref{fig:rpert_z} shows.

The change in the perturbation at high redshift is more subtle: the size of the perturbation decreases as the perturber is moved behind the lens, in both physical and angular units. The decreased size in kpc, in particular, defies the naive expectation that the larger angular diameter distance leads to a greater physical size for the perturbation radius. This is a result of the recursive lensing which gradually smears the perturber's convergence into an arc, and requires the perturber to be closer to the critical curve for an optimal fit. (A secondary effect is that the dashed line is showing the unlensed angular size of the perturbation radius, i.e. the angular size it maps to in the perturber's lens plane, which is slightly smaller than the observed angular size after being lensed by the primary lens galaxy.) Note that outside the redshift range from 0.15 to 0.35, the perturbation radius dips below 0.8 kpc, hence the projected mass within 1 kpc may be less robustly determined; on the other hand, we have seen in Figure \ref{fig:m1kpc_slope_z}(a) that the fit becomes severely degraded outside this range, so this is less of a concern. Over this redshift range, the projected mass within 1 kpc is approximately constant and greater than $2\times 10^9M_\odot$.

\end{appendix}

\end{document}